\documentstyle[prd,aps,preprint,epsf,psfig]{revtex}
\newcommand{\gsim}{ \mathop{}_{\textstyle \sim}^{\textstyle >} }
\newcommand{\lsim}{ \mathop{}_{\textstyle \sim}^{\textstyle <} }
\newcommand{\vev}[1]{ \left\langle {#1} \right\rangle }
 
\newcommand{\MEV}{~\mbox{MeV}} \newcommand{\GEV}{~\mbox{GeV}}
\newcommand{\TEV}{~\mbox{TeV}}
\def\Frac#1#2{{\displaystyle\frac{#1}{#2}}}

\begin{document}
\tighten
\draft

 \baselineskip 0.6cm
\renewcommand{\thefootnote}{\fnsymbol{footnote}}
\setcounter{footnote}{1}
\title{
\hfill{\normalsize\vbox{\hbox{TU-655}}}
~\\
\hfill{\normalsize\vbox{\hbox{UT-02-26}}}
~\\
\hfill{\normalsize\vbox{\hbox{CERN-TH/2002-098}}}
~\\
Non-thermal dark matter via Affleck--Dine baryogenesis\\
and its detection possibility}
 \vskip 1.2cm
\author{
\vskip 0.2cm Masaaki~Fujii$^{1,2}$ and K.~Hamaguchi$^{3}$}
\address{
$^1$Department of Physics, University of Tokyo, Tokyo 113-0033, Japan\\
$^2$CERN Theory Division, CH-1211 Geneva 23, Switzerland\\
$^3$Department of Physics, Tohoku University, Sendai 980-8578, Japan\\
}

\vskip 2cm
%\date{\today}
\maketitle

\vskip 2cm
\begin{abstract}
The formation and late time decays of Q-balls are generic consequences
 of the Affleck--Dine (AD) baryogenesis. A substantial amount of the
 lightest supersymmetry (SUSY) particles (LSPs) are produced
 non-thermally as the decay products of these Q-balls.  This requires a
 significantly large annihilation cross section of the LSP
so as not to overclose the
 universe, which predicts a higgsino- or wino-like LSP instead of the
 standard bino LSP.  We have reexamined the AD baryogenesis with special
 attention to the late-time decays of the Q-balls, and then specified
 the parameter regions where the LSPs produced by the Q-ball decays
 result in a cosmologically interesting mass density of dark matter by
 adopting several SUSY breaking models.  This reveals new cosmologically
 interesting parameter regions, which have not attracted much attention
 so far.  We have also investigated the prospects of direct and indirect
 detection of these dark matter candidates, and found that there is an
 intriguing possibility to detect them in various next generation dark
 matter searches.

\end{abstract}
\newpage

\renewcommand{\thefootnote}{\arabic{footnote}}
\setcounter{footnote}{0}

%%%%%%%%%%%%%%%%%%%%%%%%%%%%%%%%%%%%%%%%%%%%
%%%%%%%%%%%%%%%%%%%%%%%%%%%%%%%%%%%%%%%%%%%%
%%%%%%%%%%%%%%%%%%%%%%%%%%%%%%%%%%%%%%%%%%%%

\section{Introduction}

Both the origin and nature of dark matter and the production
mechanism of the baryon (matter--antimatter) asymmetry in the present
universe are among the most fundamental puzzles in cosmology and in
particle physics. Of the many candidates for dark matter, the
lightest supersymmetric particle (LSP) has been attracting wide
interest as an ideal candidate, which is inherent in the supersymmetric
(SUSY) standard model, and it is exactly stable under the $R$-parity
conservation.

On the other hand, once we introduce SUSY in the standard model, there
appear a number of flat directions in the scalar potential carrying
baryon ($B$) and/or lepton ($L$) number. Affleck and Dine (AD)
proposed an interesting mechanism~\cite{AD} to generate effectively a baryon
asymmetry by using these flat directions. Because of
the flatness of the potential, it is quite plausible that a linear
combination of squark and/or slepton fields has a large expectation
value along a flat direction during the inflation in the early
universe. After the end of inflation, this flat direction field
$\phi$ starts a coherent oscillation. At this stage, due to baryon
(and/or lepton) number-violating operators in the scalar potential,
the $\phi$ field gets a non-zero motion along the phase direction as
well, which means nothing but baryon (and/or lepton) number
generation since the $\phi$ carries a baryon (and/or lepton) number.

Recently, we have pointed out that the higgsino- or wino-like
neutralino naturally becomes the dominant component of the dark
matter, if either of them is the LSP and if the AD mechanism is
responsible for the generation of the observed baryon asymmetry in the
present universe~\cite{FH}.\footnote{We do not consider the
gauge-mediated SUSY breaking models~\cite{GMSB}, where the Q-ball is
generally stable~\cite{DKS,KS}.} The LSP is produced non-thermally by
the late-time decay of the Q-balls, which are generally formed in the AD
mechanism~\cite{KS,Enq-McD-PLB425,Enq-McD-NPB538}. Then, the large 
pair annihilation cross section of the 
higgsino- or wino-like LSP 
leads to the desired mass density of
dark matter, instead of overclosing the universe, as in the case of the
standard bino LSP. 
It should be noticed that the typical decay temperature of
the Q-ball is between $1\MEV$ and (a few)$\GEV$, i.e. below the freeze-out
temperature of the LSP and before the big-bang nucleosynthesis, which is
just the desired range for the present scenario.
In this scenario, it is very interesting that
both the dark matter and the baryon asymmetry
are produced by a single source, the Q-ball decay, 
which originates from the AD condensation. 
Furthermore, the
relatively strong interactions of these LSPs significantly enlarge the
possibility of detecting these dark matter 
candidates in both direct and indirect searches.

In this paper, we give more detailed analyses of this scenario,
``higgsino or wino dark matter from Q-ball decay'', adopting several
SUSY breaking models, which include the minimal supergravity (mSUGRA)
scenario, the anomaly-mediated SUSY breaking (AMSB) model~\cite{AMSB} with
additional universal scalar mass, and the no-scale model with non-universal
gaugino masses~\cite{Komine-Yamaguchi}. In each model, we will show
parameter regions in which the LSP from the Q-ball decay can be the
dominant component of the dark matter. 
This predicts new cosmologically interesting parameter regions, which 
have not attracted much attention;
the wino dark matter is realized in
wide parameter regions in the AMSB models and the no-scale models, where
the non-thermal wino production via decays of Q-balls naturally explains
the mass density of the dark matter.  Even in the mSUGRA scenario,
higgsino dark matter is realized in a relatively wide region by virtue of the
``focus point'' behaviour of $m_{H_{u}}^2$~\cite{focus-point}, in which
the mass density of dark matter is explained by the non-thermally
produced higgsino.

We also investigate the direct and indirect detection of the
neutralino dark matter in these regions. We calculate the
proton--neutralino scalar cross section for the direct detection.  As
for the indirect detection, we adopt the $\chi\chi\rightarrow
\gamma\gamma$ annihilation channel, which produces monoenergetic
$\gamma$-ray lines. We estimate the neutralino annihilation rate into
the $2\gamma$ final state using the full one-loop calculations
presented in Ref.~\cite{two-photon}, and discuss the possibility of
detection in the next generation of air Cherenkov telescopes for several
models of density profile in the halo.  Actually, the large higgsino
or wino content of the LSP strongly enhances the direct and indirect
detection rates, via the Higgs exchange diagrams and the chargino--W
boson loop diagrams, respectively.  As we will see, if the higgsino or
wino is really a significant component of dark matter, there exists an
intriguing possibility to detect these non-thermal dark matter candidates in
various next-generation detectors.

Note that, if the higgsino- or wino-like dark matter is
indeed detected at future experiments, it suggests the existence of 
non-thermal sources for these LSPs, which predicts a highly
non-trivial history of the universe. This might discriminate the
origin of the baryon asymmetry between various scenarios.

The rest of this paper is organized as follows: first, we reexamine the AD
baryogenesis scenario, giving particular attention to the late-time 
decay of Q-balls
in Section~\ref{SEC-AD}. 
This includes an interesting new twist in the AD baryogenesis 
using non-renormalizable operators in the K\"ahler potential 
to rotate the AD field. This
scenario is now able to explain the observed baryon asymmetry 
without any assumptions of additional large entropy production to
dilute the resultant baryon asymmetry, such as 
decays of heavy moduli fields and/or a thermal inflation.
In Section~\ref{SEC-Q}, we review the properties of the Q-ball in the
minimal SUSY standard model (MSSM).  We discuss
the non-thermal production of the LSP dark matter from the Q-ball decay
in Section~\ref{SEC-LSPfromQ-general}. Our main results are given in
Section~\ref{SEC-main}. We show in several SUSY breaking models the
parameter spaces where the higgsino- and wino-like LSPs from the
Q-ball decay can be the dominant component of the dark matter, and we
investigate their direct and indirect detection in detail.
We also comment on the possibility that the non-thermally produced 
neutralino via the late-time decays of Q-balls form the warm dark
matter, which erases the cuspy profile of the matter density of the 
halo that might be inconsistent with the observations.
We present concluding remarks and discussions in 
Section~\ref{conclusions}.

\section{Affleck--Dine mechanism}
\label{SEC-AD}

In this section we briefly review the Affleck--Dine
mechanism~\cite{AD,DRT} in the MSSM. Hereafter, we assume 
$R$-parity conservation and the following superpotential at
renormalizable level as usual:
\begin{eqnarray}
 W = y_U Q \bar{U} H_u + y_D Q \bar{D} H_d + y_E L \bar{E} H_d + \mu H_u H_d
  \,.
  \label{EQ-Wren}
\end{eqnarray}
Here, $Q$, $\bar{U}$, $\bar{D}$, $L$, $\bar{E}$, $H_u$ and $H_d$ denote
superfields of left-handed quark doublets, right-handed up-type and
down-type quarks, left-handed lepton doublets, right-handed charged
leptons and Higgs doublets, respectively; we omit the family indices
for simplicity. In the scalar potential, there are a number of flat
directions along which the $F$-term potential coming from the
superpotential in Eq.~(\ref{EQ-Wren}) as well as the $D$-term potential
vanish~\cite{GKM}:
\begin{eqnarray}
 && L H_u\,,H_u H_d\,,
  \nonumber \\
 &&\bar{U}\bar{D}\bar{D}\,,\, L L \bar{E}\,,\, Q\bar{D}L\,,
  \nonumber \\
 && QQQL\,,\,\bar{U}\bar{U}\bar{D}\bar{E}\,,\,
  Q\bar{U}Q\bar{D}\,,\, Q\bar{U}L\bar{E}\,,\,
  \cdots\,,
\end{eqnarray}
where the ellipsis denotes the flat directions consisting of linear
combinations with more scalar fields. In the following, we will
parametrize a flat direction by a complex scalar field $\phi$.

\subsection{Affleck--Dine mechanism with non-renormalizable operators in
the superpotential}
\label{SEC-ADwS}

In the supersymmetric limit, the flat directions are lifted only by
non-renormalizable operators in the scalar potential. Let us first assume
that all the non-renormalizable operators consistent with the MSSM gauge
symmetry and the $R$-parity exist in the superpotential. 
Most of the flat directions are then lifted by the scalar potential coming
from the following dimension-4 superpotentials:
\begin{eqnarray}
 W = \frac{\lambda}{M}L H_u L H_u\,,\, \frac{\lambda}{M}QQQL\,,\,
  \frac{\lambda}{M}\bar{U}\bar{U}\bar{D}\bar{E}\,,
\end{eqnarray}
where $M$ denotes effective scales at which these non-renormalizable
operators are induced. Here and hereafter, we discard superpotentials
that conserve the baryon and lepton numbers (e.g. $W \propto
Q\bar{U}Q\bar{D}$, $Q\bar{U}L\bar{E}$), since they cannot generate
baryon asymmetry. We have included superpotentials conserving $B-L$,
since the Q-ball can decay after the electroweak phase transition and
the baryon asymmetry is not washed out by the sphaleron effect~\cite{KRS}.

Some of remaining flat directions are lifted by the following 
dimension-6 superpotentials:
\begin{eqnarray}
 W = \frac{\lambda}{M^3}\bar{U}\bar{D}\bar{D}\,\bar{U}\bar{D}\bar{D}\,,
  \frac{\lambda}{M^3}LL\bar{E}\,LL\bar{E}\,.
\label{superpotential}
\end{eqnarray}
All the other flat directions are lifted either by $B$- and $L$-
conserving superpotentials or by superpotentials of the form  $W =
(\lambda/M^{n-3})\psi\phi^{n-1}$~\cite{GKM}. In the latter case, one
can show that $\psi$ becomes zero when the $\phi$ field develops a
large vacuum expectation value. Therefore, this type of superpotential
cannot provide a non-zero $A$-term that is indispensable to create a
net baryon/lepton asymmetry as will be discussed soon.  Thus, in the
following, we write the superpotential that lifts a flat direction
$\phi$ as
\begin{eqnarray}
 W = \frac{1}{n M^{n-3}}\phi^{n}\,,
  \label{EQ-W}
\end{eqnarray}
with $n = 4$ or $6$. Note that we have normalized the effective scale
$M$ including the coupling $\lambda$, and hence $M$ can be larger than
the cut-off scale (e.g. Planck scale).

In the case of the leptonic flat direction (e.g. $L H_u$, $LL
\bar{E}$), lepton asymmetry is first generated by the AD
mechanism. Then a part of the produced lepton asymmetry is
converted~\cite{FY} into baryon asymmetry through the sphaleron
effect.  In particular, leptogenesis via the $L H_u$ flat
direction~\cite{MY} has attracted much
attention~\cite{MM,AFHY,FHY-LHu-1,FHY-LHu-BL,FHY-LHu-0nbb,ADM}, since
in this case the baryon asymmetry in the present universe is directly
related to the neutrino mass. This allows us to obtain definite
predictions on the rate of the neutrinoless double beta
decay~\cite{FHY-LHu-1,FHY-LHu-0nbb}. In this paper, however, we do not
consider the $L H_u$ flat direction since the resultant Q-ball along
this direction is very small and evaporates well above the weak
scale~\cite{Enq-McD-PLB425,Enq-Jok-McD}. As for the other leptonic
flat directions, large Q-balls can be formed. However, the Q-balls
must decay or evaporate well before the sphaleron effect terminates
($T\sim 100\GEV$), which requires some mechanisms to make Q-balls
small enough~\cite{FHY-BL,Allahverdi-M-O}. In this case, the LSPs
produced by the decays of Q-balls are thermalized and the bino-like
LSP is the almost unique candidate for neutralino dark matter. We do
not treat such cases in the following discussions, and concentrate on
the flat directions carrying non-zero baryon number.

Let us now discuss the AD mechanism with non-renormalizable
superpotential. The relevant
scalar potential for $\phi$ is given by
\begin{eqnarray}
 V(\phi) &=& (m_\phi^2 - c_H H^2)|\phi|^2
  \nonumber\\
 &&+ \frac{m_{3/2}}{n M^{n-3}}
  \left(a_m \phi^n + h.c.\right)+
  \frac{1}{M^{2n-6}}|\phi|^{2n-2}
  \label{potential}.
\end{eqnarray}
Here, the potential terms, which are proportional to the soft mass
squared $m_\phi^2$ and the gravitino mass $m_{3/2}$, come from the SUSY
breaking at the true vacuum. We will concentrate on gravity-mediated
and gaugino-mediated SUSY breaking models, and take $m_{\phi}\simeq
m_{3/2}|a_m|\simeq 1\TEV$, hereafter. (We will discuss later the case
of anomaly-mediated SUSY breaking models, where $m_{3/2}\gg m_\phi$.)
The terms depending on the Hubble parameter $H$ denote the effect of
SUSY breaking caused by the finite energy density of the
inflaton~\cite{DRT}: $c_H$ is a real constant of order unity, which
depends on the couplings between the inflaton and the $\phi$ field in
the K\"ahler potential. Hereafter, we take $c_H\simeq 1$ ($>0$), which
is crucial to let $\phi$ have a large expectation value during 
inflation.  There might also exist a Hubble-induced $A$-term potential, which
has the same form as the second term in Eq.~(\ref{potential}), with $H$
instead of $m_{3/2}$.\footnote{ This requires the existence of a
  three-point coupling of the AD field to the inflaton in the K\"ahler
  potential, $\delta K\supset I \phi^\dag \phi/M_{pl}+{\rm h.c.}$,
  where $I$ denotes the inflaton superfield. This coupling leads to a
  relatively high reheating temperature of the inflation, $T_{R}\sim
  m_{I}(m_{I}/M_{pl})^{1/2}$, where $m_{I}$ is the mass of the
  inflaton. Therefore, as we will see, the absence of this three-point
  coupling is also desirable for obtaining the low reheating
  temperature that explains the right amount of baryon
  asymmetry. Notice that $c_H\simeq 1$ does not need this three-point
  coupling, which is consistent with the low reheating temperatures.}
Even in the presence of this term, the following discussions are not
altered,\footnote{See the related discussions in
  Refs.~\cite{DRT,MM,FHY-LHu-1}.} so we assume its absence 
in the present work, for simplicity.  The last term is the
$F$-term potential coming from the superpotential in Eq.~(\ref{EQ-W}).
Here, we assume the absence of thermal effects, which will be
justified later.

The estimation of the baryon asymmetry is rather straightforward.
The baryon number density is related to the AD field as
\begin{eqnarray}
 n_B = \beta i (\dot{\phi^*} \phi - \phi^* \dot{\phi})\;,
  \label{number-density}
\end{eqnarray}
where $\beta$ is the corresponding baryon charge of the AD field,
which is at most $1/3$.
The equation of motion for the AD field in the expanding universe is
given by
\begin{eqnarray}
 \ddot{\phi} + 3 H \dot{\phi} + \frac{\partial V}{\partial \phi^*} = 0\;.
  \label{EQ-motion}
\end{eqnarray}
Then, together with Eq.~(\ref{number-density}), the equation of
motion for the baryon number density is written as follows:
\begin{eqnarray}
\dot{n}_B+3 H n_B=&&2\beta \;{\rm Im}\left(\frac{\partial V}{\partial\phi}\phi\right)
\nonumber\\
=&&2 \beta \;
\frac{m_{3/2}}{M^{n-3}}\;{\rm Im}\left(a_{m}\phi^n\right)\,,
\label{EQ-number-density}
\end{eqnarray}
where we take $m_{3/2}$ as real by adjusting the phase of $a_{m}$.
One can see that the production rate of the baryon number is
proportional to the $A$-term.
By integrating this equation, we obtain the baryon number at the
cosmic time $t$ as
\begin{equation}
\left[R^{3}n_B\right](t)=2\beta \;\frac{|a_{m}|m_{3/2}}{M^{n-3}}\int^t
\;R^3 |\phi|^n \;{\rm sin}\theta\, dt\;,
\label{B-number}
\end{equation}
where $R$ is the scale factor of the expanding universe, and
$\theta\equiv {\rm arg}(a_{m})+n\;{\rm arg}(\phi)$.

During inflation, the negative Hubble mass term $-c_{H}H^2 |\phi|^2$
causes an instability of the flat direction field $\phi$ around the
origin, and the AD field acquires a large expectation value:
\begin{equation}
|\phi|\simeq (H_{I}M^{n-3})^{1/(n-2)}\;,
\label{initial-amp}
\end{equation}
where $H_{I}$ denotes the Hubble parameter during inflation.
This is the balance point between the negative Hubble mass term
$-c_{H}H^{2}|\phi|^2$ and the $F$-term potential
$|\phi|^{2n-2}/M^{2n-6}$. Note that the curvature along the
phase direction $m_{ph}^2$ is much smaller than $H_{I}^2$ at this point,
since
\begin{equation}
m_{ph}^2\simeq \frac{m_{3/2}}{M^{n-3}}|\phi|^{n-2}\simeq m_{3/2}H_{I}\ll
 H_{I}^2\;.
\label{curvature}
\end{equation}
Therefore, the initial phase of the AD field fixed during inflation
does not generally coincide with the minimum of the $A$-term potential in
Eq.~(\ref{potential}), and hence we naturally expect
that ${\rm sin}\theta={\cal O}(1)$.

After the end of inflation, the AD field slowly rolls down toward the
origin following the gradual decrease of the Hubble parameter $H$ as
$|\phi(t)|\simeq (H(t) M^{n-3})^{1/n-2}$. At this slow rolling regime,
the right-hand side of Eq.~(\ref{B-number}) increases as $\propto
t^{2-2/(n-2)}$.  Here, we have assumed the matter-dominated universe,
which is true as long as $T_{R}\lsim 2\times 10^{10}\GEV
\left(m_{\phi}/10^3\GEV\right)^{1/2}$. When $H(t_{\rm osc})\simeq
m_{\phi}$, the soft mass term of the AD field eventually dominates the
negative Hubble mass term, and causes the coherent oscillation of the AD
field around the origin.  After this time, the amplitude of the AD field
rapidly decreases as $|\phi|\propto t^{-1}$, and then the production of
the baryon number terminates at the time $H_{\rm osc}\simeq m_{\phi}$.

Using the above arguments and  Eq.~(\ref{B-number}), we obtain the
baryon number density at the time $t=t_{\rm osc}$:
\begin{eqnarray}
 n_B(t_{\rm osc}) = \frac{2(n-2)}{3(n-3)}
  \beta  \delta_{\rm eff}|a_m|
  m_{3/2}\left(H_{\rm osc} M^{n-3}\right)^{2/(n-2)}
  \,,
  \label{EQ-ntosc}
\end{eqnarray}
where $\delta_{\rm eff}\equiv {\rm sin}\theta (={\cal O}(1))$. Then,
after completion of the reheating process of the inflation, this
leads to the following baryon asymmetry:
\begin{eqnarray}
 \frac{n_B}{s} &=& \frac{1}{4}\frac{T_R}{M_{pl}^2 H_{\rm osc}^2}n(t_{\rm osc})
  \label{EQ-ns-tosc}
  \\
 &=&
 \frac{n-2}{6(n-3)}
  \beta  \delta_{\rm eff}|a_m|
  \frac{m_{3/2} T_R}{M_{pl}^2 H_{\rm osc}^2}
  \left(H_{\rm osc} M^{n-3}\right)^{2/(n-2)},
\end{eqnarray}
where $T_{R}$ is the reheating temperature of the inflation, $s$ is
the entropy density of the universe and $M_{\rm pl}=2.4\times 10^{18}\GEV$
is the reduced Planck scale.  By using $H_{\rm osc}\simeq m_\phi$, it
is given by
\begin{eqnarray}
 \frac{n_B}{s}\simeq 1\times 10^{-10}\times \beta  \delta_{\rm eff}|a_m|
  \left(\frac{m_{3/2}}{m_\phi}\right)
  \left(\frac{M}{M_{\rm pl}}\right)
  \left(\frac{T_R}{10^9\GEV}\right)\,,
  \label{EQ-nBs-n4}
\end{eqnarray}
for $n=4$, and
\begin{eqnarray}
 \frac{n_B}{s}\simeq 5\times 10^{-10}\times \beta  \delta_{\rm eff}|a_m|
  \left(\frac{m_{3/2}}{m_\phi}\right)
  \left(\frac{1\TEV}{m_\phi}\right)^{1/2}
  \left(\frac{M}{M_{\rm pl}}\right)^{3/2}
  \left(\frac{T_R}{100\GEV}\right)\,,
  \label{EQ-nBs-n6}
\end{eqnarray}
for $n=6$. These quantities remain
constant unless there are  additional entropy productions.

These predictions on the baryon asymmetry are not altered by the
presence of the Q-ball formation, as long as the produced Q-balls are
unstable and decay before the big-bang nucleosynthesis (BBN).
The details related to the Q-balls will be discussed in Section~\ref{SEC-Q}.
The only point remaining to be checked is the absence of the thermal effects.
In the remainder of this section, we discuss these
in the cases of the $n=4$ and $n=6$ superpotentials.
\\
\\
{ $\bf{1)\;\; n=4}$}

In the case of the $n=4$ superpotential, it was pointed out that the
thermal effects cause the early oscillations of the AD field $H_{\rm
osc}\gg m_{\phi}$, which significantly suppress the resultant baryon
asymmetry~\cite{ACE}. Analytic and more systematic investigations were
done for the case of the leptogenesis via the $LH_{u}$ flat direction,
which revealed an interesting property, the ``reheating temperature
independence'' of the baryon asymmetry~\cite{FHY-LHu-1}. For the
other flat directions, however, there are some subtleties.  The
dimension-5 operators coming from the superpotential $\propto
QQQL,\;\bar{U}\bar{U} \bar{D}\bar{E}$ are responsible for the proton
decay~\cite{Saka-Yana-Weinberg}. Null observations of the proton decay
then lead to strong constraints~\footnote{Another possibility is
the absence of these operators.  In such a case, most of the flat
directions are expected to be lifted by the $n=6$ superpotential, and 
two other operators are operative for AD baryogenesis besides
$(\bar{U}\bar{D}\bar{D})^2$, such as $\bar{U}\bar{D}\bar{D}Q\bar{D}L$,
$LL\bar{E}\bar{U}\bar{D}\bar{D}$.} on the effective cut-off scale,
$M\gsim 10^{25}\GEV$, at least for the most relevant
operators~\cite{goto-nihei,hisano-murayama}.  Following the
analyses in Ref.~\cite{FHY-LHu-1}, one can show that there is no thermal
effect as long as $T_{R}\lsim 10^{6}\GEV$ when $M\gsim 10^{23}\GEV$, and
hence there is no need to worry about the thermal effects in these
scenarios.

As we will see, such a large scale of $M$ leads to the formation of very
large Q-balls with the decay temperature $T_{d}\lsim 1\GEV$.  Since
the decays of the Q-balls occur well after the sphaleron effect
terminates, the baryon asymmetry is not washed out, although the
relevant flat directions conserve the $B-L$ symmetry. 
\\
\\
{$\bf{2)\;\;n=6}$}

In the case of the $n=6$ superpotential, it is much easier to avoid the
thermal effects because the potential is much flatter than in the 
$n=4$ case. For the appearance of the thermal mass terms, the fields
coupled with the $\phi$ field must reside in the thermal bath: 
$f|\phi|<T$, where $f$ is the Yukawa or gauge coupling constant of the
AD field.  Then, the sufficient condition to avoid the thermal mass
terms is written as
\begin{equation}
T_{R}\lsim 10^{8}\GEV\;\left(\frac{f}{10^{-5}}\right)^2\left(
\frac{M}{M_{\rm pl}}\right)^{3/2}\;.
\label{CON-thermalmass}
\end{equation}
Another thermal effect we have to check comes from the thermal
logarithmic potential~\cite{Anisimov-Dine}:
\begin{equation}
\delta V\supset a T^{4}{\rm log}\left(\frac{|\phi|^{2}}{T^{2}}\right)\;,
\label{thermal-log}
\end{equation}
where $a$ is a constant given by the fourth power of
gauge and/or Yukawa coupling constants.
To avoid the early oscillations due to this potential,
we need the following condition:
\begin{equation}
T_{R}\lsim \frac{1}{|a|^{1/2}} M \left(
\frac{m_{\phi}^{3}}{M_{\rm pl}^2 M}\right)^{1/4}=7.0\times 10^7\GEV
\left(\frac{10^{-2}}{|a|}\right)^{1/2}
\left(\frac{M}{M_{\rm pl}}\right)^{3/4}\left(\frac{m_{\phi}}{1\TEV}
\right)^{3/4}\;,
\label{CON-thermal-log}
\end{equation}
which leads to the comparable constraint on the reheating temperature
as Eq.~(\ref{CON-thermalmass}). As a result, the early oscillations
due to the thermal effects can be easily avoided as long as $T_{R}\lsim
10^8\GEV$ in AD baryo/leptogenesis with $n=6$ superpotential.

\subsection{Affleck--Dine mechanism without superpotential}
\label{SEC-ADwoS}

The other interesting possibility is
that non-renormalizable operators in the superpotential
are forbidden by some chiral symmetries, such as $R$-symmetry.
In this case, the initial amplitude of the AD field is determined
by the negative Hubble mass term $-c_{H}H^{2}|\phi|^2$ and
the non-renormalizable operators in the K\"ahler potential and is
expected to be $|\phi|_{0}\lsim M_{\rm pl}$.

In this section, as an example, we consider the following
terms~\cite{AD}, which are consistent with $R$-symmetry:
\begin{equation}
\delta {\cal L}
 =
 \int d^4 \theta
 \left(
  \lambda_{1}\frac{Z^\dag Z}{M_{\rm pl}^4}
  QQ\bar{U}^{\dag}\bar{E}^{\dag}
  +
  \lambda_{2}\frac{Z^{\dag}Z}{M_{\rm pl}^{4}}
  Q\bar{U}^{\dag}\bar{D}^{\dag}L+{\rm h.c.}
  \right)\;,
\label{Kahler}
\end{equation}
where $\lambda_{i}$'s are coupling constants and $Z$ is the superfield
with the non-vanishing $F$-term
responsible for the SUSY breaking in the true vacuum.
The effective potential of the AD field induced by these terms
and SUSY breaking effects is then given by
\begin{equation}
V=(m_{\phi}^{2}-c_{H}H^{2})|\phi|^{2}+
\frac{m_{3/2}^2}{4 M_{\rm pl}^{2}}\left(\lambda\phi^{4}+{\rm h.c.}\right)+\ldots\;,
\label{V-without-superpotential}
\end{equation}
where $\lambda$ is a coupling constant and the ellipsis denotes the
higher order terms coming from the K\"ahler potential. (Above the
Planck scale the scalar potential is expected to be exponentially
lifted by the supergravity effect.) 

The evolution of the AD field is much simpler than the case with
non-renormalizable superpotential. During  inflation,
the AD field is fixed at a very large scale $|\phi|_{0}\lsim M_{\rm pl}$
because of the negative Hubble mass term $-c_{H}H^{2}|\phi|^2$.
The AD field just stays there until $H=H_{\rm osc}\simeq m_{\phi}$
because of a large damping effect, which appears in the second term in
Eq.~(\ref{EQ-motion}).
The baryon number production completes as soon as the AD field
starts the coherent oscillation around the origin at $H=H_{\rm osc}$.
In this scenario, the AD field is completely decoupled from thermal
backgrounds, and the early oscillation of the AD field $(H_{\rm osc}
\gg m_{\phi})$ thus does not occur.

A crucial difference appearing in this scenario is the existence of a
large entropy production due to decays of the AD field. At the
reheating process of the inflation, the energy density of the inflaton is converted
into the radiation. After that, the energy density of the radiation
reduces as $\rho_{\rm rad}\propto R^{-4}$. On the other hand, the AD
field starts the coherent oscillation at $H=H_{\rm osc}$ with the
initial amplitude $|\phi|=|\phi|_{0}$, and its energy density only
reduces as $\rho_{\phi}\propto R^{-3}$. Then the AD field begins to
dominate the energy density of the universe at some time, and
substantial entropy is produced through decays of the AD field.

The AD field dominates the energy density of the universe before its
decay if the following condition is satisfied:
\begin{equation}
T_{R}> 3 T_{d}\left(\frac{M_{\rm pl}}{|\phi|_{0}}\right)^{2}\;,
\label{E-dominance}
\end{equation}
where $T_{d}$ is the decay temperature of the $\phi$ field, which
corresponds to the decay temperature of the Q-ball and much lower than
the weak scale, $T_{d}\ll m_{\rm w}$.  In Eq.~(\ref{E-dominance}), we
have assumed that the coherent oscillation of the AD field starts
before the completion of the reheating process of the inflation. If
this is not the case, i.e. if $T_R\gsim 2\times 10^{10}\GEV~
(m_\phi/1\TEV)^{1/2}$, the condition on $T_{d}$ becomes weaker than
this case.

After the decays of the AD field, the resultant baryon asymmetry is
given by the following simple form:
\begin{eqnarray}
\frac{n_{B}}{s}=&&\frac{\rho_{\phi}}{s}
\left(\frac{n_{\phi}}{\rho_{\phi}}\right)
\left(\frac{n_{B}}{n_{\phi}}\right)=\frac{3}{4}\frac{T_{d}}{m_{\phi}}\left(
\frac{n_{B}}{n_{\phi}}\right)\nonumber\\
&&=7.5\times 10^{-6}\left(
\frac{T_{d}}{10\MEV}\right)\left(\frac{1\TEV}{m_{\phi}}\right)\left(
\frac{n_{B}}{n_{\phi}}\right)\;,
\label{baryon-without-W}
\end{eqnarray}
where $n_{\phi}$ denotes the total number density of $\phi$ and
anti-$\phi$ field. The ratio of baryon to $\phi$ number is easily estimated
by an argument similar to the one given in the previous section as
\begin{equation}
  \left(
    \frac{n_{B}}{n_{\phi}}
  \right)
  \simeq
  |\lambda|
  \left(
    \frac{m_{3/2}}{m_{\phi}}
  \right)^2
  \left(
    \frac{|\phi|_{0}}{M_{\rm pl}}
  \right)^2
  \delta_{\rm eff}\;,
\label{B-phi}
\end{equation}
where $\delta_{\rm eff}\equiv {\rm sin}({\rm arg}(\lambda)+4{\rm
arg}(\phi_0))$. Then a reasonable set of parameters
\begin{equation}
  |\lambda|
  \left(
    \frac{m_{3/2}}{m_{\phi}}
  \right)^2
  \left(
    \frac{|\phi|_{0}}{M_{\rm pl}}
  \right)^2
  \delta_{\rm eff}
  \sim
  10^{-5}
  \left(\frac{10\MEV}{T_{d}}\right)
\label{EQ-condition}
\end{equation}
can naturally explain the observed baryon asymmetry.

Before the discovery of the formation of large Q-balls, decays of the AD
field have been considered to occur at $T_{d}\sim m_{\phi}$. If this is
the case, the resultant baryon asymmetry is washed out by the sphaleron
effect, since the K\"ahler potential adopted in the present work
conserves the $B-L$ symmetry.  If we adopt other interactions that violate
this symmetry,\footnote{ Such interactions consistent with
$R$-symmetry only appear at much higher orders. } the resultant ratio of 
$\phi$-number to entropy is given by $n_{\phi}/s\sim {\cal O}(1)$,
which requires a very small ratio of baryon to $\phi$-number 
$n_{B}/n_{\phi}\sim 10^{-10}$. Such a small ratio results in a very
unnatural fine tuning on $\lambda$ and/or $\delta_{\rm eff}$, and hence
additional large entropy productions have been considered to be
necessary after the decays of the AD field, such as a thermal
inflation or decays of heavy moduli fields.

Interestingly enough, the large initial amplitude of the AD field
results in the formation of very large Q-balls with low decay
temperature $T_{d}={\cal O}(10~{\rm MeV})$.\footnote{See
  Sec.~\ref{SEC-Q}.} This significantly enhances the entropy
production via decays of the AD fields and naturally explains the
observed baryon asymmetry.  Higgsino or wino LSP is the necessary
condition for this scenario to work, so as not to overclose the
universe and explain the right amount of dark matter.

A quite beautiful point in this scenario is that the resultant baryon
asymmetry and dark matter density are solely determined by the sector
related to the AD field and annihilation cross section of the LSP. Any
other process cannot affect the final result as long as the AD
fields (stored in the Q-balls) dominate the energy density of the
universe at their decay time. Especially, the final baryon asymmetry
is completely independent of the reheating temperature of the
inflation.

In addition, the so-called ``cosmological gravitino
problem''~\cite{grav-prb} is also solved.  The dilution factor, which is
the ratio of the entropy density before and after the decays of the AD
field, is given by
\begin{equation}
\frac{1}{\Delta}\equiv \frac{s_{\rm before}}{s_{\rm after}}=
3\frac{T_{d}}{T_{\rm ini}}\left(\frac{M_{\rm pl}}{|\phi|_{0}}\right)^{2},
\label{dilution-fac}
\end{equation}
where
\begin{equation}
T_{\rm ini}={\rm Min}\left[
T_{R},\quad\sqrt{m_{\phi} M_{\rm pl}}\left(\frac{90}{\pi^2 g_{*}}\right)^{1/4}
\right]\;,
\label{def-Tini}
\end{equation}
with $g_*$ being the effective degrees of freedom at temperature
$T$. It can be seen from these relations that even if the reheating
temperature of the inflation is much higher than $\sim 10^{12}\GEV$,
the produced gravitinos are sufficiently diluted for there not to be any
gravitino problem.

\subsection{Affleck--Dine mechanism in anomaly-mediated SUSY breaking models}
\label{SEC-ADinAMSB}

In anomaly-mediated SUSY breaking (AMSB) models, SUSY is broken in a
hidden sector, and it is transmitted to the MSSM sector dominantly via
the super-Weyl anomaly. The MSSM gauginos and matter fields obtain
soft masses, which are one-loop-suppressed relative to the gravitino
mass. Therefore, the order of the gravitino mass is estimated as
$m_{3/2}\sim m_{\rm soft}/\alpha\gg m_{\rm soft}$.

A crucial point is that the AD mechanism always uses the
non-renormalizable operators to violate baryon and/or lepton number,
which also violate the super-Weyl symmetry at tree level. Hence, the SUSY
breaking effects induced in these operators are of the order of the
gravitino mass and not loop-suppressed, which generates a global
minimum for the AD field displaced from the
origin~\cite{K-Watari-Y}.

First consider the AD mechanism using non-renormalizable operators in
the superpotential. In this case, the scalar potential of the AD field
is given by Eq.~(\ref{potential}), but with much larger gravitino mass.
During the inflation, the negative Hubble mass term drives the AD
field far from the origin and fixes it at $|\phi|\simeq
(H_{I}M^{n-3})^{1/n-2}$. After the inflation ends, the amplitude of
the AD field gradually decreases as $|\phi|\simeq (H
M^{n-3})^{1/n-2}$. At this stage, the curvature along the phase
direction $m_{ph}^2$ is given by $m_{ph}^2\simeq m_{3/2} H$. Thus,
after $H\lsim m_{3/2}$, this large curvature forces the AD field to
settle down on  the bottom of one of the valleys of the $A$-term potential,
which are located at
\begin{equation}
{\rm arg}(a_{m})+n \;{\rm arg}(\phi)=\pi,\;\;{\rm mod}\;\pi\,.
\label{valley-inAMSB}
\end{equation}
Consequently, the AD field is inevitably trapped at the global minimum
on the way toward the origin along the bottom of the valley:
\begin{equation}
  |\phi|_{\rm min}\simeq \left(m_{3/2}M^{n-3}\right)^{1/(n-2)}\;.
  \label{global-minimum}
\end{equation}
This results in an SU(3)$_{C}$ and/or U(1)$_{\rm EM}$ breaking
universe.  To avoid this disaster, we need some mechanisms to stop the
AD field during the inflation below ``the top of the hill'' in the
scalar potential, which is located at
\begin{equation}
|\phi|_{0}\lsim |\phi|_{\rm hill}\simeq \left(\frac{m_{\phi}^2}{m_{3/2}}M^{n-3}\right)^{1/(n-2)}\;.
\label{ADinAMSBwithsuper}
\end{equation}

The same difficulty appears in the AD mechanism making use of the
non-renormalizable operators in the K\"ahler potential.  In this case,
the potential is written as in Eq.~(\ref{V-without-superpotential}),
with larger gravitino mass $m_{3/2}\gg m_\phi$, and the global minimum
is located near the Planck scale.  In order to avoid a colour and/or
U(1)$_{\rm EM}$ breaking universe, we need to impose the following
condition:
\begin{equation}
|\phi|_{0}\lsim |\phi|_{\rm hill}=M_{\rm pl}\frac{m_{\phi}}{m_{3/2}}\;.
\label{ADinAMSBwithoutsuper}
\end{equation}

One natural solution is to gauge the U(1)$_{B-L}$
symmetry~\cite{FHY-BL}.\footnote{In Ref.~\cite{FHY-BL}, we make
  use of the gauged $B-L$ symmetry to make the Q-ball small enough for
  the produced bino LSPs  not to overclose the universe. In the
  present work, we concentrate on the case of the higgsino or wino LSP,
  so that there is no need for such an adjustment on the $B-L$ breaking
  scale. } If the $B-L$ symmetry is gauged, flat directions along
which the AD fields cannot cancel the U(1)$_{B-L}$ $D$-term
contributions within themselves can be lifted at the $B-L$ breaking
scale $v$ by the $D$-term potential.  Then, the amplitude of the AD
field can be easily suppressed by adjusting the $B-L$ breaking scale
as $v\lsim |\phi|_{\rm hill}$.  Even if we suppress the initial
amplitude of the AD field, there remains  plenty of room to explain
the observed baryon asymmetry and to form large Q-balls, which allows
us to obtain the right amount of wino dark matter in AMSB models.

%%%%%%%%%%%%%%%%%%%%%%%%%%%%%%%%%%%%%%%%%%%%%%%%%%%%%%%%
\section{Q-ball}
\label{SEC-Q}
%%%%%%%%%%%%%%%%%%%%%%%%%%%%%%%%%%%%%%%%%%%%%%%%%%%%%%%%

After the flat direction field $\phi$ starts its coherent oscillation, a
kind of non-topological soliton ``Q-ball''~\cite{Q-ball} is formed
because of
spatial instabilities of the $\phi$
field~\cite{KS,Enq-McD-PLB425,Enq-McD-NPB538}. In this section we
briefly review the formation and the decay of the Q-ball.

\subsection{Size of the Q-ball}

The relevant scalar potential of $\phi$ at the time of Q-ball
formation is given by
\begin{eqnarray}
 V(\phi) = m_\phi^2
  \left(1 + K \log \left(\frac{|\phi|^2}{M_G^2}\right)\right)
  |\phi|^2\;,
\label{potential-Q-ball-formation}
\end{eqnarray}
where $M_G$ is the renormalization scale at which the soft mass $m_\phi$
is defined, and the $K\log(|\phi|^2)$ term represents the one-loop
correction. This mainly comes from the gaugino loops and $K$
is estimated in the range from $-0.01$ to
$-0.1$~\cite{Enq-McD-PLB425,Enq-McD-NPB538,Enq-Jok-McD}. 
Because of 
the potential flatter than $\phi^2$, spatial instabilities of the
homogeneous $\phi$ field are caused after it starts the coherent
oscillation~\cite{KS}. In the momentum space, the instability band is
given by~\cite{Kasu-Kawa-PRD62}
\begin{eqnarray}
 0 < \frac{k^2}{R^2} < 3 m_\phi^2 |K|\,,
\end{eqnarray}
where $k$ is the comoving momentum of the fluctuations of the $\phi$
field. The best amplified mode is given by the centre of the band:
$(k^2/R^2)_{\max}\simeq (3/2)m_\phi^2|K|$. This corresponds to the
radius of the Q-ball, which is estimated analytically using the
Gaussian profile of the Q-ball~\cite{Enq-McD-NPB538}:
\begin{eqnarray}
 R_Q\simeq \frac{1}{m_\phi}\sqrt{\frac{2}{|K|}}\,.
  \label{EQ-RQ}
\end{eqnarray}
For the best amplified mode, the perturbations $\delta \phi$ of the
$\phi$ field grow according to the following equation:
\begin{eqnarray}
 \left|\frac{\delta \phi}{\phi}\right|
  &\simeq&
  \left|\frac{\delta \phi}{\phi}\right|_{\rm osc}
  \exp\left[\int dt \frac{3 m_\phi |K|}{4}\right]
  \nonumber\\
 &=&
  \left|\frac{\delta \phi}{\phi}\right|_{\rm osc}
  \exp\left[\frac{m_\phi |K|}{2 H}\right]\,,
\end{eqnarray}
where we have used $H^{-1} = (3/2)t$, assuming that the Q-ball formation
occurs before the reheating process of the inflation is
completed.\footnote{In the AD mechanism without non-renormalizable
superpotentials, the reheating temperature can be as high as $T_R\gsim
2\times 10^{10}\GEV (m_\phi/1\TEV)^{1/2}$ and the Q-ball formation might
occur after the reheating phase. (See Sec.~\ref{SEC-ADwoS}.) In this
case, $H^{-1}=2t$, and Eq.~(\ref{EQ-ntnon}) is modified as $n_B(t_{\rm
non})\simeq n_B(t_{\rm osc})\times (H_{\rm non}/H_{\rm osc})^{3/2}$. In
such a case, the charge of the Q-ball becomes larger by a factor of
${\cal O}(10)$.} Therefore, the fluctuation becomes non-linear when the
Hubble parameter becomes
\begin{eqnarray}
 H = H_{\rm non} = \frac{m_\phi |K|}{2 \alpha}\,,
  \label{EQ-Hnon}
\end{eqnarray}
where $\alpha \equiv \ln (|\phi|_{\rm osc}/|\delta \phi|_{\rm osc})$ and
$|\delta\phi|_{\rm osc}\approx (2\pi R_Q)^{-1}$ for the best amplified
mode~\cite{Enq-McD-NPB570}. For the parameters we adopt in the present
analyses, its typical value is given by $\alpha\simeq 30$--$40$.  At
this time ($H = H_{\rm non}$), the baryon number density of the AD
condensation becomes
\begin{eqnarray}
 n_B(t_{\rm non})\simeq n_B(t_{\rm osc})\times
  \left(\frac{H_{\rm non}}{H_{\rm osc}}\right)^2\,.
  \label{EQ-ntnon}
\end{eqnarray}
Thus, from Eqs.~(\ref{EQ-RQ}), (\ref{EQ-Hnon}) and (\ref{EQ-ntnon}) and
$H_{\rm osc}\simeq m_\phi$, the typical charge of a single Q-ball is
estimated to
\begin{eqnarray}
 Q &\simeq& \frac{4}{3}\pi R_Q^3\times n_B(t_{\rm non})
  \nonumber\\
  &\simeq&
  \frac{2\sqrt{2}\pi}{3}
  \frac{|K|^{1/2}}{\alpha^2}
  \frac{1}{m_\phi^3}n_B(t_{\rm osc})
  \nonumber\\
 &\simeq&
  \overline{\beta}_a
  \left(\frac{|\phi_{\rm osc}|}{m_\phi}\right)^2
  \epsilon
  \,,
\end{eqnarray}
where
\begin{eqnarray}
 \overline{\beta}_a\equiv
  6\times 10^{-4}
  \left(\Frac{30}{\alpha}\right)^2
  \left(\Frac{|K|}{0.03}\right)^{1/2}
  \,,
  \qquad
  \epsilon\equiv
  \frac{n_B(t_{\rm osc})}{n_\phi(t_{\rm osc})}\;(\le\beta)
  \,.
\end{eqnarray}
Here, we have used $n_\phi(t_{\rm osc})\simeq m_\phi |\phi|_{\rm
  osc}^2$. This estimation of the Q-ball charge is roughly consistent
with the result obtained by numerical lattice
simulations~\cite{Kasu-Kawa-PRD62,Kasu-Kawa-PRD64},
\begin{eqnarray}
 Q\simeq
  \overline{\beta}_n
  \left(\frac{|\phi_{\rm osc}|}{m_\phi}\right)^2
  \times
  \left\{
  \begin{array}{ll}
   \epsilon & {\rm for}\quad \epsilon\gsim \epsilon_c\,,\\
   \epsilon_c & {\rm for}\quad \epsilon\lsim \epsilon_c\,,
  \end{array}
  \right.
\end{eqnarray}
where
\begin{equation}
 \overline{\beta}_n
  \equiv
  6\times 10^{-3}
  \,,
  \qquad
  \epsilon_c \sim 0.01
  \,.
\end{equation}
Notice that for $\epsilon \lsim \epsilon_c$ the Q-ball charge becomes
independent of the initial charge density $\epsilon$. This is because
negative charge Q-balls are created in this
region~\cite{Kasu-Kawa-PRD62}.  In the following discussions, we shall
adopt the following equation in estimating the Q-ball charge:
\begin{eqnarray}
 Q=Q_{\rm max}\times
   \left\{
   \begin{array}{ll}
    \Frac{\epsilon}{\beta} & {\rm for}\quad\epsilon_{c}\lsim\epsilon\le
 \beta\,,\\
\\
    \Frac{\epsilon_c}{\beta} & {\rm for}\quad \epsilon\lsim \epsilon_c\,,
   \end{array}
  \right.
  \label{EQ-Q-tosc}
\end{eqnarray}
where 
\begin{equation}
Q_{\rm max}\sim 3\times 10^{-3}\left(\frac{|\phi_{\rm osc}|}{m_{\phi}}\right)^2\beta\;.
\label{Q-max}
\end{equation}

Let us now estimate the charge of the Q-ball in each of the cases discussed in
the previous sections. In the case of AD baryogenesis with
non-renormalizable superpotential (Sec.~\ref{SEC-ADwS}), the Q-ball
charge is estimated by using Eqs.~(\ref{EQ-ntosc}) and
(\ref{EQ-Q-tosc}), resulting in
\begin{eqnarray}
 \epsilon \simeq
  \frac{2(n-2)}{3(n-3)}
  \beta  \delta_{\rm eff}|a_m|
  \left(\frac{m_{3/2}}{m_\phi}\right)
  \;,
\label{epsiron}
\end{eqnarray}
\begin{eqnarray}
 Q &\sim&
  3\times10^{-3}
  \times
  \frac{2(n-2)}{3(n-3)}
  \beta  \delta_{\rm eff}|a_m|
  \left(\frac{m_{3/2}}{m_\phi}\right)
  \left(\frac{M}{m_\phi}\right)^{2(n-3)/(n-2)}
  \nonumber\\
 &\simeq&
  \left\{
   \begin{array}{lcc}
    3\times 10^{20}\times \beta\delta_{\rm eff} |a_m|
     \left(\Frac{m_{3/2}}{m_\phi}\right)
     \left(\Frac{1\TEV}{m_\phi}\right)
     \left(\Frac{M}{10^{26}\GEV}\right)
     & \quad{\rm for} & n=4\,,
     \\
    3\times 10^{20}\times \beta\delta_{\rm eff} |a_m|
     \left(\Frac{m_{3/2}}{m_\phi}\right)
     \left(\Frac{1\TEV}{m_\phi}\right)^{3/2}
     \left(\Frac{M}{M_{\rm pl}}\right)^{3/2}
     & \quad{\rm for} & n=6\,.
   \end{array}
   \right.
\end{eqnarray}

As mentioned in Section~\ref{SEC-ADwoS}, the Q-ball charge becomes larger
in the AD mechanism without non-renormalizable superpotential, since the
initial amplitude of the $\phi$'s oscillation is large. Recall that a
relatively smaller value of $\epsilon = n_B/n_\phi < \epsilon_c$ is
preferred in order to generate the correct amount of baryon asymmetry
[see Eqs.~(\ref{B-phi}) and (\ref{EQ-condition})]. Thus, from
 Eq.~(\ref{EQ-Q-tosc}), the Q-ball charge is estimated to be
\begin{eqnarray}
 Q \sim 10^{28}
  \left(\frac{|\phi|_0}{M_{\rm pl}}\right)^2
  \left(\frac{1\TEV}{m_\phi}\right)^2
  \epsilon_c
  \,,
\end{eqnarray}
which results in $Q\sim 10^{24}$--$10^{26}$ for $|\phi|_0\simeq
(0.1$--$1)\times M_{\rm pl}$.

Finally, in the case of the AMSB model, the initial amplitude of the AD
field $\phi$ should be suppressed to avoid the unwanted trapping of the
AD field at the global minimum, as discussed in Section~\ref{SEC-ADinAMSB}.
{}From Eqs.~(\ref{ADinAMSBwithsuper}), (\ref{ADinAMSBwithoutsuper}) and
(\ref{EQ-Q-tosc}), the Q-ball charge is estimated to
\begin{eqnarray}
 Q &\lsim&  10^{-3}
  \left(
  \frac{M^{n-3}}{m_{3/2}m_\phi^{n-4}}
  \right)^{2/(n-2)}
   \nonumber\\
 &\simeq&
  \left\{
   \begin{array}{lcc}
    3\times 10^{18}\times
     \left(\Frac{30\TEV}{m_{3/2}}\right)
     \left(\Frac{M}{10^{26}\GEV}\right)
     & \quad{\rm for} & n=4\,,
     \\
    2\times 10^{19}\times
     \left(\Frac{30\TEV}{m_{3/2}}\right)^{1/2}
     \left(\Frac{1\TEV}{m_\phi}\right)
     \left(\Frac{M}{M_{\rm pl}}\right)^{3/2}
     & \quad{\rm for} & n=6\,,
   \end{array}
   \right.
\end{eqnarray}
for the case with non-renormalizable superpotentials, and
\begin{eqnarray}
 Q &\lsim& 10^{-3}
  \left(
  \frac{M_{\rm pl}}{m_{3/2}}
  \right)^2
   \nonumber\\
 &\simeq&
  6\times 10^{24}
  \left(\frac{30\TEV}{m_{3/2}}\right)^2
  \,,
\end{eqnarray}
for the case without non-renormalizable superpotentials.

\subsection{Decay of the Q-ball}

Before discussing the decay of the Q-ball, we comment on its 
evaporation~\cite{Lain-Shap,Banej-Jedam}. Because of the large
expectation value of the scalar field inside the Q-ball, most of it
is decoupled from the surrounding thermal bath. However, a thin
outer region of the  Q-ball is thermalized, since particles in the thermal
plasma can penetrate into this region, and hence a partial evaporation
of the Q-ball charge occurs. As stressed in Ref.~\cite{Banej-Jedam}, the
evaporation of the charge from the Q-ball surface is suppressed by
diffusion effects, and the evaporation occurs most effectively at $T
\sim m_\phi$. The total amount of the evaporated charge is estimated
to
\begin{eqnarray}
 \Delta Q \sim 10^{18}
  \times
  \left(\frac{0.03}{|K|}\right)^{1/2}
  \left(\frac{1\TEV}{m_\phi}\right)
  \,,
\end{eqnarray}
for $T_R\gsim m_\phi$, and
\begin{eqnarray}
 \Delta Q \sim 10^{16}
  \times
  \left(\frac{0.03}{|K|}\right)^{1/2}
  \left(\frac{1\TEV}{m_\phi}\right)^3
  \left(\frac{T_R}{100\GEV}\right)^2
  \,,
\end{eqnarray}
for $T_R\lsim m_\phi$. (Here, we have used $T\simeq (T_R^2 M_{\rm pl}
H)^{1/4}$ before the reheating process of the inflation is completed.)
Therefore, as long as the initial charge of the Q-ball is larger than
${\cal O}(10^{18})$, most part of the Q-ball charge survives the
evaporation. Hereafter, we assume $Q\gsim {\cal O}(10^{18})\GEV$ (which
is naturally realized in the present scenarios, as shown in the previous
sections) and discuss the emission of the remaining charge by the decay
of the Q-ball.

The remaining charge of the Q-ball is emitted through its decay 
into light fermions. The decay rate is estimated as~\cite{CCGM}
\begin{eqnarray}
 \Gamma_Q \equiv -\frac{d Q}{d t}
  \lsim
  \frac{\omega^3 {\cal A}}{192 \pi^2}
  \,,
  \label{EQ-Q-decayrate}
\end{eqnarray}
where ${\cal A} = 4\pi R_Q^2$ is the surface area of the Q-ball and
$\omega\simeq m_\phi$. This upper bound is likely to be saturated for
$\phi(0)\gg m_\phi$, where $\phi(0)$ is the field value of the AD field
at the centre of the Q-ball~\cite{CCGM}.\footnote{If kinematically
allowed, there are also decay channels into lighter scalars through
3-point couplings. Even if this is the case, the decay rate is at most
comparable to that in Eq.~(\ref{EQ-Q-decayrate}).}  Therefore, the
lifetime of a single Q-ball with an initial charge $Q_i$ is estimated 
to be
\begin{eqnarray}
 \tau_d
  =
  \frac{Q_i}{\Gamma_Q}
  &\gsim&
  1\times 10^{-7}~{\rm sec}
  \times
  \left(\frac{|K|}{0.03}\right)
  \left(\frac{1\TEV}{m_\phi}\right)
  \left(\frac{Q_i}{10^{20}}\right)
  \,,
\end{eqnarray}
or equivalently, the decay temperature $T_d$ of the Q-ball is given by
\begin{eqnarray}
 T_d
  &\lsim&
  2 \GEV
  \times
  \left(\frac{0.03}{|K|}\right)^{1/2}
  \left(\frac{m_\phi}{1\TEV}\right)^{1/2}
  \left(\frac{10^{20}}{Q_i}\right)^{1/2}
  \,.
\end{eqnarray}
Therefore, the Q-ball decays at $T_d\sim 1\MEV$--(a few)$\GEV$ for the
Q-ball charge estimated in the previous sections. It is quite
interesting to observe that the Q-ball decay occurs just after the
freeze-out of the LSP ($T_f\sim m_\chi/20$) and just before the
beginning of the big-bang nucleosynthesis (BBN) ($T\sim 1\MEV$). Thus,
the Q-ball decay can naturally provide both the baryon asymmetry
required from the BBN and the non-thermal source of the LSP. It
should also be noticed that the decay occurs after the electroweak phase
transition ($T\sim 100\GEV$). Thus, the produced baryon asymmetry is not
washed out by the sphaleron effect, even if the $B-L$ is conserved in the
AD mechanism, as in the $n=4$ cases and the $W = 0$ cases.

\section{non-thermal dark matter from the Q-ball decay}
\label{SEC-LSPfromQ-general}

We now turn to the discuss of the non-thermal production of the LSP from the
Q-ball decay. 
The relation between the Q-ball number density and baryon number density 
is given by
\begin{eqnarray}
&&n_{B}=Q_{i}(n_{Q}^{+}-n_{Q}^{-})\;,\nonumber\\
\nonumber\\
&&n_{Q}^{\rm total}\equiv n_{Q}^{+}+n_{Q}^{-}=\frac{n_{B}}{Q_{i}}
\times \left\{
\begin{array}{ll}
    1 & {\rm for}\quad\epsilon_{c}\lsim\epsilon\le
 \beta\,,\\
\\
    \Frac{\epsilon_c}{\epsilon} & {\rm for}\quad \epsilon\lsim \epsilon_c\,,
   \end{array}
  \right.
\label{nQ}
\end{eqnarray}
where $Q_{i}$ is the absolute value of the initial charge of a
single Q-ball, which is expected to be of the same order for
positive  and negative charge Q-balls, and $n_{Q}^+$
($n_{Q}^-$) is the number density of positive (negative) charge Q-balls. 
Here, the negative charge Q-balls only appear in the case
of $\epsilon\lsim \epsilon_{c}$. Notice that
almost all the baryon asymmetry is initially stored in the
Q-balls~\cite{Kasu-Kawa-PRD62}. 
%
%Let
%us denote the fraction of the total baryon asymmetry which is
%initially contained in the Q-balls by $f$. Then the number density
%$n_Q$ and the initial charge $Q_i$ of the Q-ball are related to the
%baryon number density $n_B$ as
%\begin{eqnarray}
% Q_i\,n_Q &=& f\,n_B\,
%  \theta(Q_i - \Gamma_Q t)
%  \nonumber\\
% &=&
%  f\,s \left(\frac{n_B}{s}\right)_0
%  \theta(Q_i - \Gamma_Q t)
%  \,,
%\end{eqnarray}
%where we have normalized the $n_B$ by the present baryon-to-entropy
%ratio $(n_B/s)_0$, since we assume that there is no entropy production
%at the Q-ball decay. The theta function represents the fact that the
%Q-balls disappear for $t \ge \tau_d = Q_i/\Gamma_Q$.

First we assume that the Q-ball decay does not produce a
large extra entropy. As can be seen in Eqs.~(\ref{EQ-nBs-n4}) and
(\ref{EQ-nBs-n6}), the AD mechanism with non-renormalizable
superpotentials can naturally provide the empirical baryon asymmetry
without any extra entropy production, if the reheating temperature of
the inflation is relatively low (e.g. $T_R\sim 100\GEV$ for the $n=4$ case
with $M \sim 10^{26}\GEV$, and for the $n=6$ case with $M
\sim M_{\rm pl}$.)  In these cases, the number density of the Q-ball is
directly related to the baryon asymmetry in the present universe.
It is easy to see that the
energy density $\rho_Q$ of the Q-ball is much less than that of the
radiation $\rho_{\rm rad}$ for $T > T_d$:
\begin{eqnarray}
 \frac{\rho_Q}{\rho_{\rm rad}}
  &\simeq&
  \frac{\rho_{\phi}}{\rho_{\rm rad}}
  \nonumber\\
 &=&
  \frac{3}{4}\epsilon^{-1}\frac{m_\phi}{T}
  \left(\frac{n_B}{s}\right)_0
  \nonumber \\
 &\lsim&
  10^{-4}\times
  \epsilon^{-1}
  \left(\frac{m_\phi}{1\TEV}\right)
  \left(\frac{n_B/s|_0}{10^{-10}}\right)
\left(\frac{1\MEV}{T}\right)
  \ll 1\qquad
  {\rm for}
  \quad T > T_d \gsim 1\MEV\,.
\end{eqnarray}
Therefore, the assumption of no large extra entropy
production at the Q-ball decay is justified for $\epsilon>10^{-4}\times(1\MEV/T_{d})$.
{}From Eq.~(\ref{epsiron}), we can expect that $\epsilon={\cal O}(0.1)$ when 
we use non-renormalizable operators in the superpotential 
to lift the flat direction.

The production rate of the LSP per unit time per unit volume is given by
\begin{eqnarray}
 &&\left(
  \frac{d n_\chi}{d t}
  \right)_{\rm prod}
  =
  N_\chi \Gamma_Q n_Q^{\rm total}
  =
  N_\chi
  \,s \left(\frac{n_B}{s}\right)_0
  \frac{\theta(\tau_d - t)}{\tau_d}\times f(\epsilon)
  \nonumber\\
\nonumber\\
&&\qquad\qquad\qquad f(\epsilon)\equiv\left\{
\begin{array}{ll}
    1 & {\rm for}\quad\epsilon_{c}\lsim\epsilon\le
 \beta\,,\\
\\
    \Frac{\epsilon_c}{\epsilon} & {\rm for}\quad \epsilon\lsim \epsilon_c\,,
   \end{array}
  \right.
\end{eqnarray}
where $N_\chi$ is the number of LSPs produced per baryon number, which
is at least $N_{\chi}^{\rm min}=(\epsilon f(\epsilon))^{-1}\ge 3$. 

The evolution of the number density of the LSP is then
described by the following Boltzmann equation:
\begin{eqnarray}
 \dot{n_\chi} + 3 H n_\chi
 =
 N_\chi \,s
 \left(
 \frac{n_B}{s}
 \right)_0
 \,
 \frac{\theta(\tau_d - t)}{\tau_d}f(\epsilon)
 -
 \vev{\sigma v}n_\chi^2\,,
 \label{EQ-Beq-0}
\end{eqnarray}
where $\vev{\sigma v}$ is the thermally averaged annihilation cross
section of the LSP.\footnote{The LSPs are likely to be in kinetic
  equilibrium at least for $T\gsim {\cal O}({\rm
    MeV})$~\cite{neutralino-WDM}.} Here, we have neglected the effect
of the pair production of the LSPs, which is suppressed by a Boltzmann
factor $\exp(-m_\chi/T)$ for $T < m_\chi$.

If the Q-ball decay produces significant entropy, such as the AD
mechanism without superpotential as discussed in Section~\ref{SEC-ADwoS},
$n_Q$ is not directly related to the present baryon asymmetry. In this
case, the evolution of the number density of the LSPs $n_\chi$ is governed by
the following coupled equations:
\begin{eqnarray}
 \dot{n_\chi} + 3 H n_\chi
  &=&
  N_\chi \Gamma_Q n_Q^{\rm total} - \vev{\sigma v} n_\chi^2\,,
  \label{EQ-Beq-1}
\end{eqnarray}
where $n_Q^{\rm total}$ is given by
\begin{eqnarray}
 \dot{n_Q}^{\rm total} + 3 H n_Q^{\rm total} = 0 & \quad ({\rm for} \quad t \le \tau_d)\,,&
 \nonumber\\
 n_Q^{\rm total} = 0 & \quad ({\rm for} \quad t > \tau_d)\,,&
 \label{EQ-Beq-2}
\end{eqnarray}
and the Hubble parameter $H$ is obtained from the Friedman equation:
\begin{eqnarray}
 H^2 &=& \frac{1}{3 M_{pl}^2}
  \left(\rho_Q + \rho_\chi + \rho_{\rm rad}\right)
  \,,
  \label{EQ-Beq-3}
\end{eqnarray}
where
\begin{eqnarray}
 \rho_Q = \rho_{\phi}&=&(\epsilon f(\epsilon))^{-1} m_\phi \left(Q_i - \Gamma_Q t\right) n_Q^{\rm total}
  \,,
  \label{EQ-Beq-4}
  \\
\rho_\chi &=&m_\chi n_\chi
  \label{EQ-Beq-5}
  \\
  \dot{\rho}_{\rm rad} + 4 H \rho_{\rm rad}
  &=&
  \left((\epsilon f(\epsilon))^{-1} m_\phi - N_\chi m_\chi \right)\Gamma_Q n_Q^{\rm total}
  +
  m_\chi \vev{\sigma v}n_\chi^2
  \,.
  \label{EQ-Beq-6}
\end{eqnarray}

Although the Boltzmann equations have complicated forms, the final
abundance of the LSPs can be approximately expressed by a simple
analytical form. Note that the Boltzmann equations for the LSP
[Eqs.~(\ref{EQ-Beq-0}) and (\ref{EQ-Beq-1})] are reduced to the
following one for $t > \tau_d$:
\begin{eqnarray}
 \dot{n_\chi} + 3 H n_\chi = -\vev{\sigma v}n_\chi^2
  \qquad
  ({\rm for}\quad t > \tau_d)
  \,.
\end{eqnarray}
It may be useful to rewrite this equation in terms of $Y_\chi =
n_\chi/s$ and the temperature $T$:
\begin{eqnarray}
 \frac{dY_\chi}{dT}
  =
  \sqrt{\frac{8\pi^2 g_*}{45}}
  \left(1 + \frac{T}{g_*}\frac{d g_*}{dT}\right)
  \vev{\sigma v}
  M_{\rm pl}
  Y_\chi^2
  \,.
\end{eqnarray}
We assume that the $s$-wave contributions 
dominate the annihilation cross section of the neutralino,
which is a reasonable approximation for $\widetilde{H}$- and
$\widetilde{W}$-like LSP with $T_{d}\ll m_{\chi}$.
By using the approximations $g_*(T)\simeq g_*(T_d)\simeq {\rm const}$ and
$\vev{\sigma v}(T)\simeq {\rm const}$, it can be solved analytically:
\begin{eqnarray}
 Y_{\chi}(T)
  \simeq
  \left[
   \frac{1}{Y_{\chi}(T_d)}
   +
  \sqrt{
   \frac{8\pi^2 g_*(T_d)}{45}
   }
   \vev{\sigma v}
    M_{\rm pl}
    (T_d - T)
   \right]^{-1}
   \,.
   \label{EQ-analytic-YT}
\end{eqnarray}

If initial abundance $Y_{\chi}(T_d)$ is large enough, the final
abundance $Y_{\chi 0}$ for $T \ll T_d$ is given by
\begin{eqnarray}
 Y_{\chi 0}
  \simeq Y_{\chi}^{\rm approx}\equiv
  \left[
   \sqrt{
   \frac{8\pi^2 g_*(T_d)}{45}
   }
   \vev{\sigma v}
   M_{\rm pl}
   T_d
   \right]^{-1}
   \,.
   \label{EQ-Ychi-analytic}
\end{eqnarray}
Therefore, in this case, 
the final abundance $Y_{\chi 0}$ is determined only by the
Q-ball decay temperature $T_d$ and the annihilation cross section of
the LSP $\vev{\sigma v}$, independently of the initial value
$Y_{\chi}(T_d)$ as long as $Y_{\chi}(T_d)\gg Y_{\chi}^{\rm approx}$.
In terms of the density parameter $\Omega_\chi$, it is
rewritten as
\begin{eqnarray}
 \Omega_{\chi}
 &\simeq&
  0.5
  \left(\frac{0.7}{h}\right)^2
  \times
  \left(
   \frac{m_{\chi}}{100 \GEV}
   \right)
   \left(
    \frac{10^{-7}\GEV^2}{\vev{\sigma v}}
    \right)
    \times
    \left(
     \frac{100 \MEV}{T_d}
     \right)
     \left(
      \frac{10}{g_*(T_d)}
      \right)^{1/2}
      \,,
      \label{Omega-ana}
\end{eqnarray}
where $h$ is the present Hubble parameter in units of $100\,\, {\rm
  km}\,\, {\rm sec}^{-1} {\rm Mpc}^{-1}$ and $\Omega_{\chi}\equiv
\rho_{\chi}/\rho_c$. ($\rho_{\chi}$ and $\rho_c$ are the energy
density of the LSP and the critical energy density in the present
universe, respectively.)

In the case of $Y_{\chi}(T_{d})<Y_{\chi}^{\rm approx}$, the final
abundance is given by 
\begin{equation}
Y_{\chi 0}
\simeq Y_{\chi}(T_{d})\gsim \epsilon^{-1}\left(\frac{n_{B}}{s}
\right)_{0}\; . 
\end{equation}
This is the case for the LSP whose annihilation cross section is small
enough, such as the bino-like LSP.  In this case, there is a very
interesting feature that the relic abundance of the LSPs is directly
related to the observed baryon asymmetry~\cite{Enq-McD-NPB538}.
Unfortunately, however, the relics of the bino-like neutralino
overclose the universe unless we assume an extremely light bino. This
is easily seen from the following density parameter:
\begin{equation}
  \Omega_{\chi}\ge \epsilon^{-1} \left(\frac{m_{\chi}}{m_{p}}
  \right)\Omega_{B}\,,
\end{equation}
where $m_{p}$ is the mass of the  nucleon.
Therefore, the bino mass should be lighter than
\begin{equation}
  m_{\chi} \le 1 {\rm ~GeV}\times 
  \left(\frac{\epsilon}{0.1}\right)
  \left(\frac{\Omega_\chi}{10\,\Omega_B}\right)\,.
\end{equation}
%unless we assume an extremely light bino, which is 
%already experimentally excluded. This is easily seen  from the 
%following density parameter:
%\begin{equation}
%\Omega_{\chi}\gsim \epsilon^{-1} \left(\frac{m_{\chi}}{m_{p}}
%\right)\Omega_{B}=10^3 \left(\frac{\epsilon^{-1}}{10}\right)
%\left(\frac{m_{\chi}}{100\GEV}\right)\Omega_{B}\;,
%\end{equation}
%where $m_{p}$ is the mass of the  nucleon and 
%$\Omega_{B}h^2=(0.004$--$0.023)$~\cite{baryonkana-}.

\begin{figure}[t!]%%%%%%%%%%%%%%%%%%%%%%%%%%%%%%%%%%%%%%%%%%%
%%%%%%%%%%%%%%%%%%%%%%%%%%%%%%%%%%%%%%%%%%%%%%%%%%%%%%%%%%%%
 \centerline{\psfig{figure=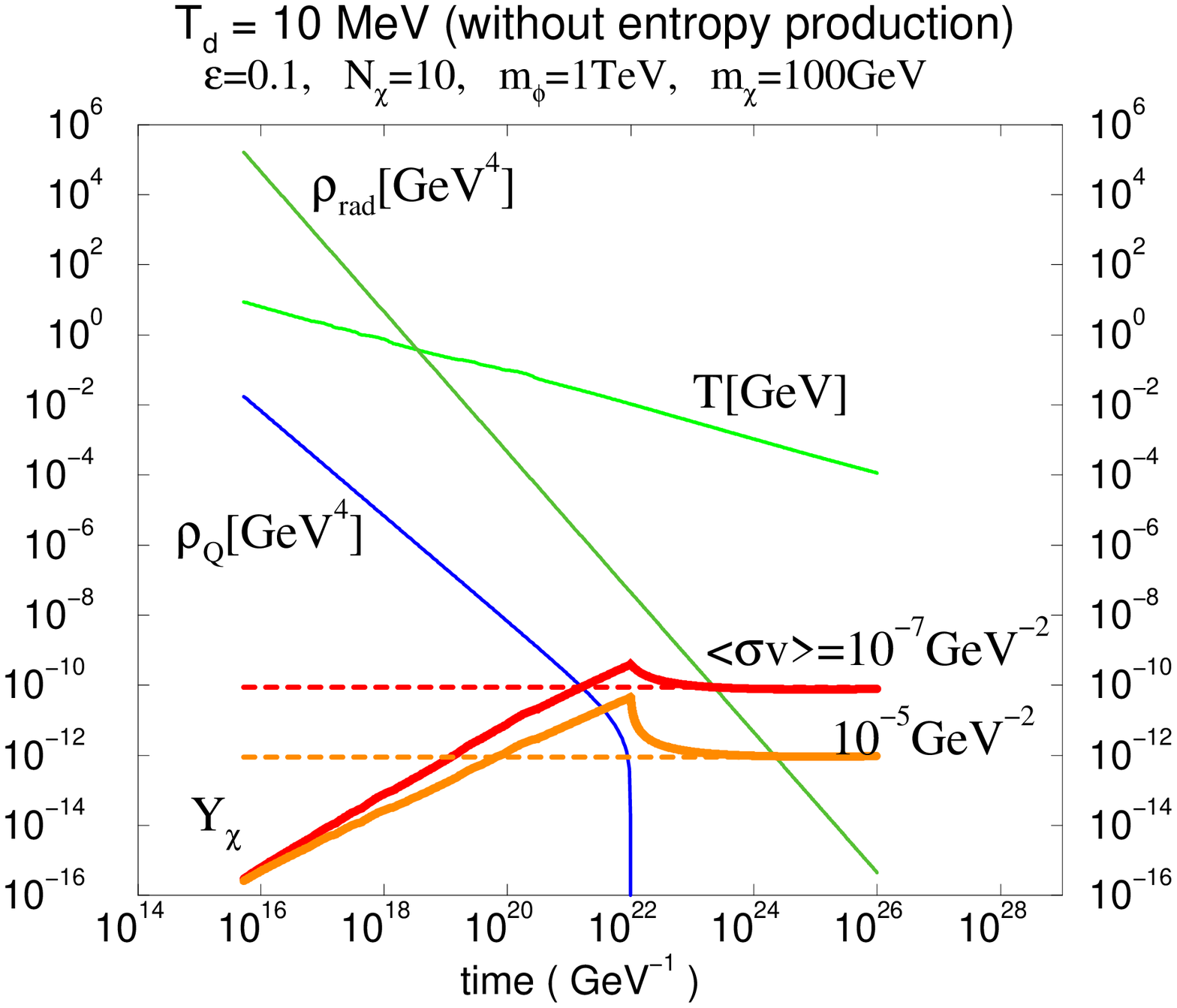,height=8cm}} 
 \vspace{-0.4cm}
 \caption{The evolution of the abundance of the neutralino dark matter
 generated from the Q-ball decay for $T_d = 10\MEV$ with $\vev{\sigma v}
 = 10^{-7}\GEV^{-2}$ and $10^{-5}\GEV^{-2}$, which are represented by
 thick solid lines. The abundances estimated by the analytic formula in
 Eq.~(\ref{EQ-Ychi-analytic}) are shown in dashed lines. In this figure,
 we have assumed that the energy density of the Q-ball is small enough
 with respect to that of the radiation. The parameters are taken to be
 $m_\phi = 1\TEV$, $m_\chi = 100\GEV$, $\epsilon = 0.1$ and $N_\chi=
 10$.} \label{FIG-Boltzmann1}
%%%%%%%%%%%%%%%%%%%%%%%%%%%%%%%%%%%%%%%%%%%%%%%%%%%%%%%%%%%%
\end{figure}

\begin{figure}[h!]%%%%%%%%%%%%%%%%%%%%%%%%%%%%%%%%%%%%%%%%%%%
%%%%%%%%%%%%%%%%%%%%%%%%%%%%%%%%%%%%%%%%%%%%%%%%%%%%%%%%%%%%
 \centerline{\psfig{figure=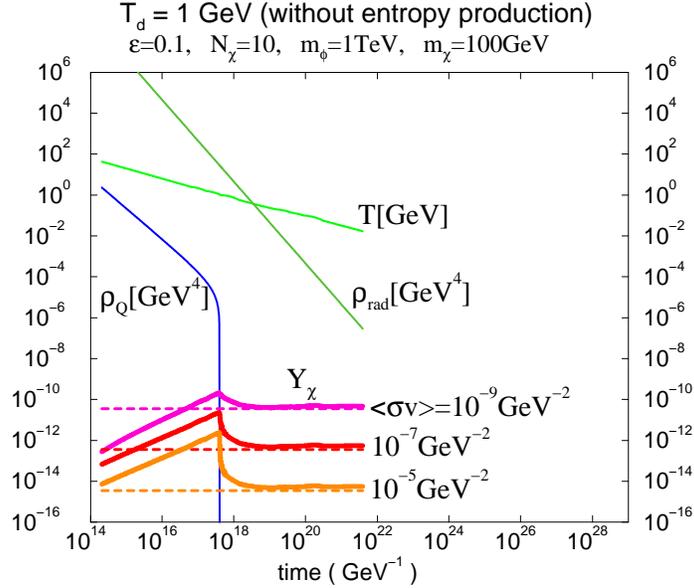,height=8cm}}
 \vspace{-0.4cm}
 \caption{The same as Fig.~\ref{FIG-Boltzmann1}, but with $T_d = 1\GEV$
 and $\vev{\sigma v}=10^{-9}$, $10^{-7}$, and $10^{-5}\GEV^{-2}$.}
 \label{FIG-Boltzmann2}
%%%%%%%%%%%%%%%%%%%%%%%%%%%%%%%%%%%%%%%%%%%%%%%%%%%%%%%%%%%%
\end{figure}

\begin{figure}[t!]%%%%%%%%%%%%%%%%%%%%%%%%%%%%%%%%%%%%%%%%%%%
%%%%%%%%%%%%%%%%%%%%%%%%%%%%%%%%%%%%%%%%%%%%%%%%%%%%%%%%%%%%
 \centerline{\psfig{figure=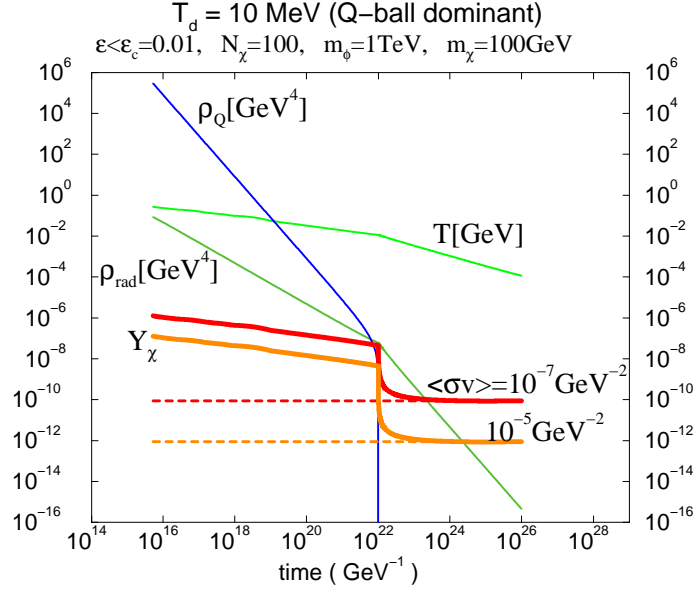,height=8cm}} 
 \vspace{-0.2cm}
 \caption{The evolution of the abundance of the neutralino dark matter
 generated from the Q-ball decay for $T_d = 10\MEV$ with $\vev{\sigma v}
 = 10^{-7}\GEV^{-2}$ and $10^{-5}\GEV^{-2}$, 
which are represented by thick solid
 lines. The abundances estimated by the analytic formula in
 Eq.~(\ref{EQ-Ychi-analytic}) are shown in dashed lines. In this figure,
 we have assumed that the Q-balls dominate the energy density of the
 universe before their decay. The parameters are taken to be $m_\phi =
 1\TEV$, $m_\chi = 100\GEV$, $\epsilon < \epsilon_c = 0.01$ and $N_\chi=
 100$.}  \label{FIG-Boltzmann3}
%%%%%%%%%%%%%%%%%%%%%%%%%%%%%%%%%%%%%%%%%%%%%%%%%%%%%%%%%%%%
\end{figure}

\begin{figure}[h!]%%%%%%%%%%%%%%%%%%%%%%%%%%%%%%%%%%%%%%%%%%%
%%%%%%%%%%%%%%%%%%%%%%%%%%%%%%%%%%%%%%%%%%%%%%%%%%%%%%%%%%%%
 \centerline{\psfig{figure=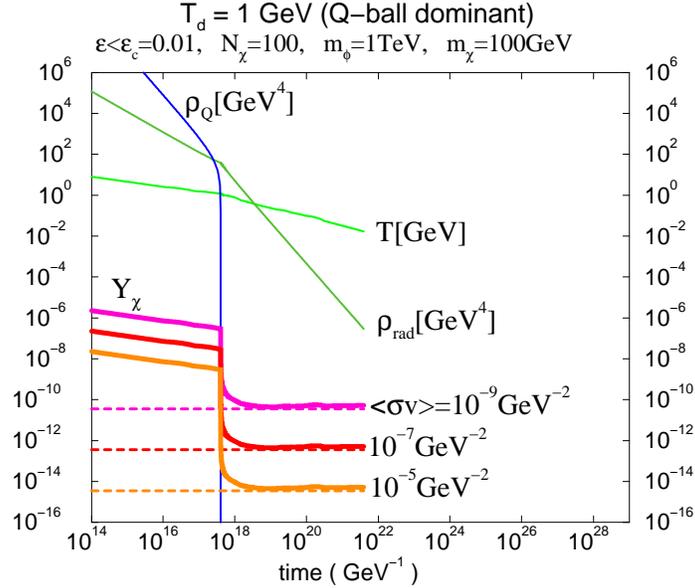,height=8cm}}
 \vspace{-0.2cm}
 \caption{The same as Fig.~\ref{FIG-Boltzmann3}, but with $T_d = 1\GEV$
 and $\vev{\sigma v}=10^{-9}$, $10^{-7}$, and $10^{-5}\GEV^{-2}$.}
\label{FIG-Boltzmann4}
%%%%%%%%%%%%%%%%%%%%%%%%%%%%%%%%%%%%%%%%%%%%%%%%%%%%%%%%%%%%
\end{figure}

We have numerically solved the Boltzmann equations for the case without 
a large entropy production [Eq.~(\ref{EQ-Beq-0})] and for the case with  
a large entropy production [Eqs.~(\ref{EQ-Beq-1})--(\ref{EQ-Beq-6})].
Here, we have used the following set of parameters:
$N_\chi = 10$, $\epsilon=0.1$, $(n_B/s)_0 = 0.7\times 10^{-10}$,
$m_{\phi}=1\TEV$ and $m_{\chi}=100\GEV$ for the former case, 
and $N_\chi = 100$, $\epsilon<\epsilon_{c}=0.01$, 
$m_{\phi}=1\TEV$ and $m_{\chi}=100\GEV$ for the latter case.
As can be seen from 
Figs.~\ref{FIG-Boltzmann1}--\ref{FIG-Boltzmann4}, the numerical calculations 
reproduce the simple analytic estimation in 
Eq.~(\ref{EQ-Ychi-analytic}) quite well.
Therefore, we use the analytic estimation in
Eqs.~(\ref{EQ-Ychi-analytic})
and (\ref{Omega-ana})
for the relic abundance of
the neutralinos in the remainder of this paper, since it gives us the
correct relic abundance of the LSPs as long as it is
the required mass density as dark matter.

Before closing this section, we should comment on the distribution of
the LSPs. So far, we have assumed that they  are  uniformly
distributed after they are produced by the Q-ball decay. One might wonder if
the LSPs are localized near the Q-ball, since the Q-ball is a
localized object. If this is the case, the pair annihilation rate of
the LSP is enhanced and the final abundance might become
smaller. Here, we show that this is not the case. As can be seen in
Eq.~(\ref{EQ-analytic-YT}), the abundance of LSPs approaches its
final value only after $(T_d-T)/T_d\simeq {\cal O}(1)$, which means it
takes a time scale $\Delta t\sim \tau_d$. (This is independent of the
local abundance, as long as it is large enough.) By that time, LSPs
have spread out from the decaying Q-ball by a random walk colliding
with background particles. Then the LSPs produced from a single
Q-ball form a Gaussian distribution around that Q-ball, whose radius
is given by $\bar{r}\simeq \sqrt{\nu\tau_d}$, where $\nu^{-1}\simeq
G_F^2 m_\chi T_d^4$~\cite{Enq-McD-NPB538}. ($G_F$ is the Fermi
constant.)  The number of the Q-balls within this radius is given by
$(4\pi/3)\bar{r}^3n_Q\sim 10^{10}\times
(T_d/1\GEV)^{-6}(Q_i/10^{20})^{-1}(m_\chi/100\GEV)^{-3/2}$, which is
much larger than 1. Therefore, the assumption of an uniform
distribution is justified.

\section{Direct and indirect detections of higgsino and
wino non-thermal dark matter from
Affleck--Dine baryogenesis}
\label{SEC-main}

In this section, we investigate the prospects of direct and indirect
detection of non-thermal dark matter resulting from the late-time
decays of Q-balls in several SUSY breaking models.  The requirement
that the LSPs produced via the late-time decays of Q-balls do not
overclose the universe leads to the parameter region where the
annihilation cross section of the LSP is substantially large. The
possible LSP candidates are higgsino $\widetilde{H}$ and wino
$\widetilde{W}$.\footnote{Here, we do not mean the pure higgsino 
  and wino LSP.
  A significant fraction of the bino component is possible.  In fact,
  this is the case for most of the parameter regions in the mSUGRA
  scenario we will present in this section.  In the case of large
  $\tan\beta$, even the bino-like LSP is possible to achieve the
  desired mass density of dark matter.}

%\footnote{Another interesting possibility is 
%the case of the sneutrino LSP. The late-time decays of Q-balls 
%are expected to provide the desired mass  density of 
%dark matter also in this case. The detailed investigation of this 
%case will be given elsewhere~\cite{F-H-sneutrino}.}

The most promising way to confirm the existence of neutralino dark
matter is given by a direct detection through elastic scatterings of
neutralino with matter.  The interactions of neutralino with matter
are usually dominated by scalar couplings for relatively heavy nuclei
$A\gsim 20$~\cite{scalar-spin,SUSYDM}.  These interactions are
mediated by the light $h$ and heavy $H$ Higgs exchanges, or a sfermion
$\tilde{f}$ exchange. Notice that the former interactions contain
$h\chi\chi$ and $H\chi\chi$ couplings, which are suppressed for
bino-like LSPs.  In the case of the wino-like dark matter, these
couplings are enhanced by the factor $g2/(g1 \tan\theta_W)$.  As for
the higgsino-like dark matter (except for the case of the pure
higgsino LSP), these couplings are highly enhanced by the mixing with
the gaugino components in the LSP. As we will see, these effects give
us an intriguing possibility to detect these non-thermal dark matter
in next generation direct dark matter searches, such as
CDMS~\cite{CDMS}, CRESST~\cite{CRESST}, EDELWEISS II~\cite{EDELWEISS},
GENIUS~\cite{GENIUS} and ZEPLIN MAX~\cite{ZEPLIN}.
%which are suppressed for
%bino-like LSPs.  As we will see, these Higgs exchange interactions
%give large direct detection rates for $\widetilde{H}$ and
%$\widetilde{W}$ dark matter, which are within the reach of the next
%generation direct dark matter searches, 

Many other indirect detection methods open up if a significant portion
of halo dark matter consists of the neutralino LSP.  These indirect
methods utilize the fact that neutralinos accumulated in the halo or
in massive bodies such as the Earth or the Sun may annihilate,
resulting in ordinary particles, which can be detected.  For example,
fluxes of antiprotons and positrons, which are not produced in large
quantities by cosmic rays, can be enhanced by the annihilation of
neutralinos accumulated in the halo.  Although the fluxes could be a
measurable size, a clear discrimination from the background is
difficult, because of the rather featureless spatial and energy
distributions of these antiparticles.  A much better signature is
provided by neutralino annihilation into neutrinos near the centre of
the Sun or the Earth. Since the ordinary solar neutrinos only have
energies at most of the order of MeV, a multi-GeV neutrino signal from
the Sun and the Earth may give us unmistakable evidence, although
there are always isotropic and anisotropic backgrounds coming from
neutrinos created on the other side of the Earth and in the outer
region of the Sun by cosmic rays, respectively.

In the present work, we investigate another promising way of detecting
neutralinos in the halo.  An excellent signature is provided by
neutralino annihilation into the $2\gamma$ final state through
one-loop diagrams~\cite{two-photon}.\footnote{ Neutralino annihilation
  into the $Z\gamma$ final state has a similar signature. For a
  detailed discussion of the indirect detections utilizing the
  neutralino annihilation channel including a photon final state, see
  Ref.~\cite{z-photon}. } This annihilation channel leads to
monoenergetic $\gamma$-rays with energy $\simeq m_{\chi}$. Although
there is an extragalactic $\gamma$ ray background and backgrounds of
gamma-like hadronic and electron showers, the signal will stand out
against them in favourable circumstances.  In fact, this is the case for
the present scenario.  As we will see, a large fraction of
$\widetilde{H}$ or $\widetilde{W}$ component in the LSP required from
the late-time Q-ball decays significantly enhances the annihilation
cross section via diagrams including a chargino--W boson loop, which
gives us an intriguing possibility to observe the $\gamma$-ray lines
in the next generation of air Cherenkov telescopes observing the
galactic centre, such as VERITAS~\cite{VERITAS}, HESS~\cite{HESS} and
MAGIC~\cite{MAGIC}.

In this section, we specify the parameter region where the neutralino
LSPs, produced non-thermally through late-time decays of Q-balls, give
a desired mass density of dark matter. In the estimation of the
annihilation cross section of the neutralino, we have used all the
tree-level annihilation channels with non-zero $s$-wave
contributions~\cite{Nojiri-1,SUSYDM}.  Here, we have neglected the
possible co-annihilation effects with the lightest chargino, which is
justified as long as the decay temperature of Q-balls is lower than
the mass difference between the LSP and the lightest chargino.  As we
will see, this condition is satisfied in most of the parameter region.
Anyway, we are not interested in the difference by a factor of ${\cal
  O}(1)$ in the estimation of the relic abundance, since there are
ambiguities in the estimation of the decay temperature of Q-balls.

We also perform calculations of
the neutralino--proton scalar cross section $\sigma_{P}$ and
annihilation cross section into the $2\gamma$ final states
$\sigma_{2\gamma}$ in several SUSY breaking models: these include
the ``focus point''~\cite{focus-point} in a mSUGRA scenario, the anomaly-mediated
SUSY breaking model~\cite{AMSB} with additional universal soft scalar
masses, and the no-scale model with non-universal gaugino
masses~\cite{Komine-Yamaguchi}. Finally, we comment on the possibility
that the non-thermally produced LSPs via Q-ball decays form the warm dark matter.

\subsection{Parameter space and possibility of detection in a  mSUGRA scenario}

In the framework of minimal supergravity (mSUGRA),
there are four continuous free parameters and one binary choice:
\begin{equation}
m_{0},\; M_{1/2},\;A_{0},\;{\rm tan}\beta,\; {\rm sign}(\mu)\;,
\label{paramter}
\end{equation}
where $m_{0},\;M_{1/2},\;A_{0}$ are the universal soft scalar mass,
gaugino mass, and trilinear scalar coupling given at the GUT scale
$M_{G}\simeq 2\times 10^{16}\GEV$, respectively. All the couplings and
mass parameters at the weak scale are obtained through the
renormalization group (RG) evolution. In our work, we use the SOFTSUSY
code~\cite{SOFTSUSY} to calculate quantities at the weak scale, which
include two-loop RG equations, one-loop self-energies for all the
particles and one-loop threshold corrections from SUSY particles to
the gauge and Yukawa coupling constants following the methods of
Ref.~\cite{precision-susy}.

At the weak scale, the higgsino mass parameter $|\mu|$
is determined by the condition of electroweak symmetry breaking,
which, at tree level, is given by
\begin{equation}
\frac{1}{2}m_{Z}^2=\frac{m_{H_{d}}^2-m_{H_{u}}^2 {\rm tan}^2\beta}{{\rm tan}^2\beta-1}-\mu^2\simeq -m_{H_{u}}^2-\mu^2\;,
\label{condition-EWSB}
\end{equation}
where the last relation holds for moderate and large ${\rm
tan}\beta\gsim 5$.

Unfortunately,
in a mSUGRA scenario, the bino-like LSP is realized in most of the
parameter space where even the thermal relics of the bino LSPs overclose the
universe. In such a region, of course, the 
bino LSPs non-thermally produced through decays of Q-balls only make the problem worse.

However, there exist an interesting region where the
$\widetilde{H}$-like LSP is naturally realized.  The crucial
observation is the existence of a ``focus point'' behaviour in the
weak scale value of the soft scalar mass of the up-type Higgs
multiplet, $m_{H_{u}}^2$~\cite{focus-point}. The weak scale value of
$m_{H_{u}}^2$ remains the weak scale even for multi-TeV $m_{0}$, as
long as $M_{1/2}$ and $A_{0}$ are set to be around the weak
scale. Therefore, from the relation in Eq.~(\ref{condition-EWSB}), one
can see that the electroweak scale is insensitive to the input
parameters for relatively large ${\rm tan}\beta$.  By virtue of this
focus point behaviour, multi-TeV values of $m_{0}$ do not require
strong fine tunings on the input parameters and are natural in that
sense. Such large values of $m_{0}$ give positive contributions to
$m_{H_{u}}^2$, making it less negative. This, in turn, leads to
smaller values of $|\mu|$ at the weak scale.  Therefore, as $m_{0}$
increases further, the $\widetilde{H}$ content in the lightest
neutralino $\chi$ increases, until it finally enters the region
$|\mu|\lsim 105\GEV$, which is excluded by chargino searches
at LEP II~\cite{chargino-search}.

In the focus point region, the large $\widetilde{H}$ content of the LSP
enhances the neutralino annihilation cross section into the W bosons via
chargino exchange, which is not helicity-suppressed. This naturally
results in a desired mass density of dark matter via the late-time
decays of Q-balls.

\begin{figure}[t!]%%%%%%%%%%%%%%%%%%%%%%%%%%%%%%%%%%%%%%%%%%%
%%%%%%%%%%%%%%%%%%%%%%%%%%%%%%%%%%%%%%%%%%%%%%%%%%%%%%%%%%%%
 \centerline{\psfig{figure=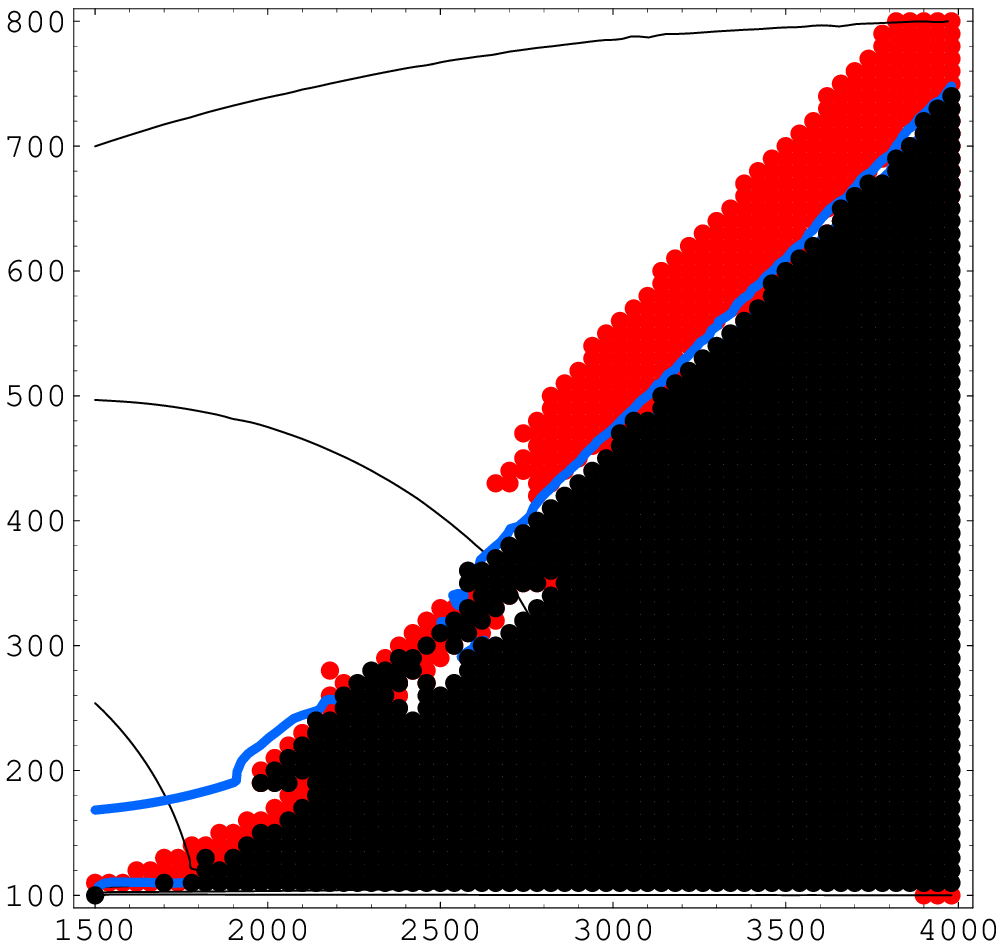,height=10cm}}
 %%%%%%%%%%%%%%%%%%%%%%%%%
 \begin{picture}(0,0)%%%%%
 %%%%%%%%%%%%%%%%%%%%%%%%
  \put(55,306){$M_{1/2}$[GeV]}  
   \put(200,1){$m_{0}$[GeV]}
\normalsize
 %%%%%%%%%%%%%%%%%%%%%%%%%
 \end{picture}%%%%%%%%%%%%
 %%%%%%%%%%%%%%%%%%%%%%%%%
\vspace{0.5cm}
 \caption{The allowed region in the mSUGRA scenario with ${\rm
tan}\beta=15$ and $A_{0}=0$ in the $(m_{0}$--$M_{1/2})$ plane.  In the red shaded region, non-thermally produced LSPs
via decays of Q-balls result in a cosmologically interesting mass
density.  The black shaded region is where the electroweak symmetry
breaking cannot be implemented. The region below the blue (thick) line
is excluded by the chargino mass bound $m_{\chi^{\pm}}\gsim 105\GEV$.
The contours of the light Higgs boson mass are given by the black (thin)
lines, which correspond to $m_{h}=117,\;120,\;122\GEV$,
respectively. There is no region excluded by the bounds on the
$b\rightarrow s\gamma$ branching ratio. }
\label{FIG-forcus15}
%%%%%%%%%%%%%%%%%%%%%%%%%%%%%%%%%%%%%%%%%%%%%%%%%%%%%%%%%%%%
\end{figure}

In Fig.~\ref{FIG-forcus15}, we show the allowed region for
${\rm tan}\beta=15$ in the $(m_{0}$--$M_{1/2})$ plane.
Here we take the sign of $\mu$ to be positive and $A_{0}=0$. 
As for the criterion to
select the region, we have imposed the following conditions:
\begin{eqnarray}
0.05\le\Omega_{\chi}h^2\le 0.5\;,\nonumber\\
1\MEV\le T_{d}\le 10\GEV\;.
\label{criterion}
\end{eqnarray}
We will use these conditions throughout this paper.  

In the red shaded
region, the non-thermally produced LSPs result in a cosmologically
interesting mass density. In the allowed region, the content of
the LSP is dominated by $\widetilde{H}$.
A typical decay temperature of Q-balls, 
which leads to the desired mass density of dark matter, 
is $100\MEV \lsim T_{d}\lsim ({\mbox{a few}})\GEV$. 
On the other hand, we have confirmed 
that the mass difference between the $\widetilde{H}$-like LSP and the 
lightest chargino is at least ${\cal O}(10)\GEV$. Hence, the
co-annihilation effects are safely neglected. 
This is also true for the case of ${\rm tan}\beta=40$, which will be 
 discussed later.
The electroweak symmetry breaking does not
occur in the black shaded region. The region below the blue (thick) line is
excluded by the chargino mass limit, $m_{\chi^\pm}\gsim105\GEV$.
The black (thin) lines are the contours of the light Higgs boson mass,
which are $117,\;120$ and $122\GEV$, respectively.
We also calculate the branching ratio of the  $b\rightarrow s\gamma$
transition~\cite{bsg-bound}. We adopt the following constraints on the
$b\rightarrow s\gamma$ branching ratio obtained by the CLEO experiment~\cite{CLEO}:
\begin{equation}
2\times 10^{-4}< B(B\rightarrow X_{s}\gamma)<4\times 10^{-4}\;.
\label{BSG}
\end{equation}
In Fig.~\ref{FIG-forcus15}, there is no region excluded by these bounds.

\begin{figure}[t!]%%%%%%%%%%%%%%%%%%%%%%%%%%%%%%%%%%%%%%%%%%%
%%%%%%%%%%%%%%%%%%%%%%%%%%%%%%%%%%%%%%%%%%%%%%%%%%%%%%%%%%%%
 \centerline{\psfig{figure=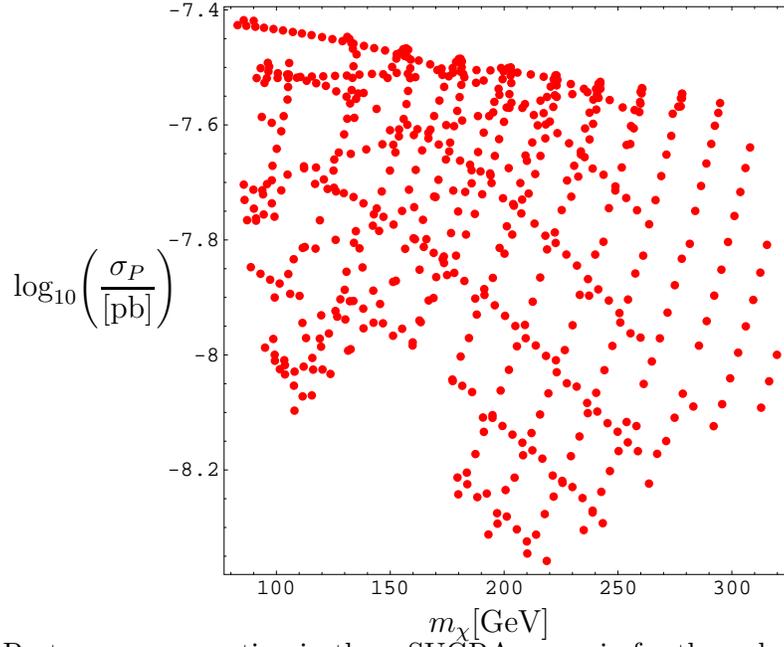,height=8cm}}
 %%%%%%%%%%%%%%%%%%%%%%%%%
 \begin{picture}(0,0)%%%%%
 %%%%%%%%%%%%%%%%%%%%%%%%%
  \put(43,130){${\rm log}_{10}\displaystyle{
\left(\frac{\sigma_{P}}{\rm [pb]}\right)}$}  
  \put(200,3){$m_{\chi}$[GeV]}
\normalsize
 %%%%%%%%%%%%%%%%%%%%%%%%%
 \end{picture}%%%%%%%%%%%%
 %%%%%%%%%%%%%%%%%%%%%%%%%
\vspace{0.cm}
 \caption{Proton--$\chi$ cross section in the mSUGRA scenario for the red
region in Fig.~\ref{FIG-forcus15}, also with the chargino mass bound.}
\label{FIG-direct-forcus15}
%%%%%%%%%%%%%%%%%%%%%%%%%%%%%%%%%%%%%%%%%%%%%%%%%%%%%%%%%%%%
\end{figure}

In general, the matter--$\chi$ cross sections are dominated by
spin-independent couplings for relatively heavy nuclei, $A\gsim
20$. Assuming this is the case, we can obtain the detection rate of the
neutralinos for each detector material by scaling the proton--$\chi$
cross section. In Fig.~\ref{FIG-direct-forcus15}, we show 
this cross section in pb units in a mSUGRA scenario with
${\rm tan}\beta=15$. Each point in this scattered plot
corresponds to one parameter set in the
red shaded region above the chargino mass bound in
Fig.~\ref{FIG-forcus15}.
In this calculation, we
adopt the following values of the proton matrix elements for each of
the three light quarks:
\begin{equation}
f_{Tu}=0.019,\quad f_{Td}=0.041,\quad f_{Ts}=0.14,
\end{equation}
where $f_{Tq}\equiv\left< p|m_{q}\bar{q}q|p\right>/m_{p}$.
For details about the calculation of the proton--$\chi$ cross section, see
Refs.~\cite{Nojiri-2,SUSYDM}. One can see from the figure that the
proton--$\chi$ cross section satisfies $\sigma_{P}\gsim 10^{-8}{\rm
pb}$ in most of the parameter space, which is within the reach
of various next-generation 
detectors~\cite{CDMS,CRESST,EDELWEISS,GENIUS,ZEPLIN}.

\begin{figure}[h!]%%%%%%%%%%%%%%%%%%%%%%%%%%%%%%%%%%%%%%%%%%%
%%%%%%%%%%%%%%%%%%%%%%%%%%%%%%%%%%%%%%%%%%%%%%%%%%%%%%%%%%%%
 \centerline{\psfig{figure=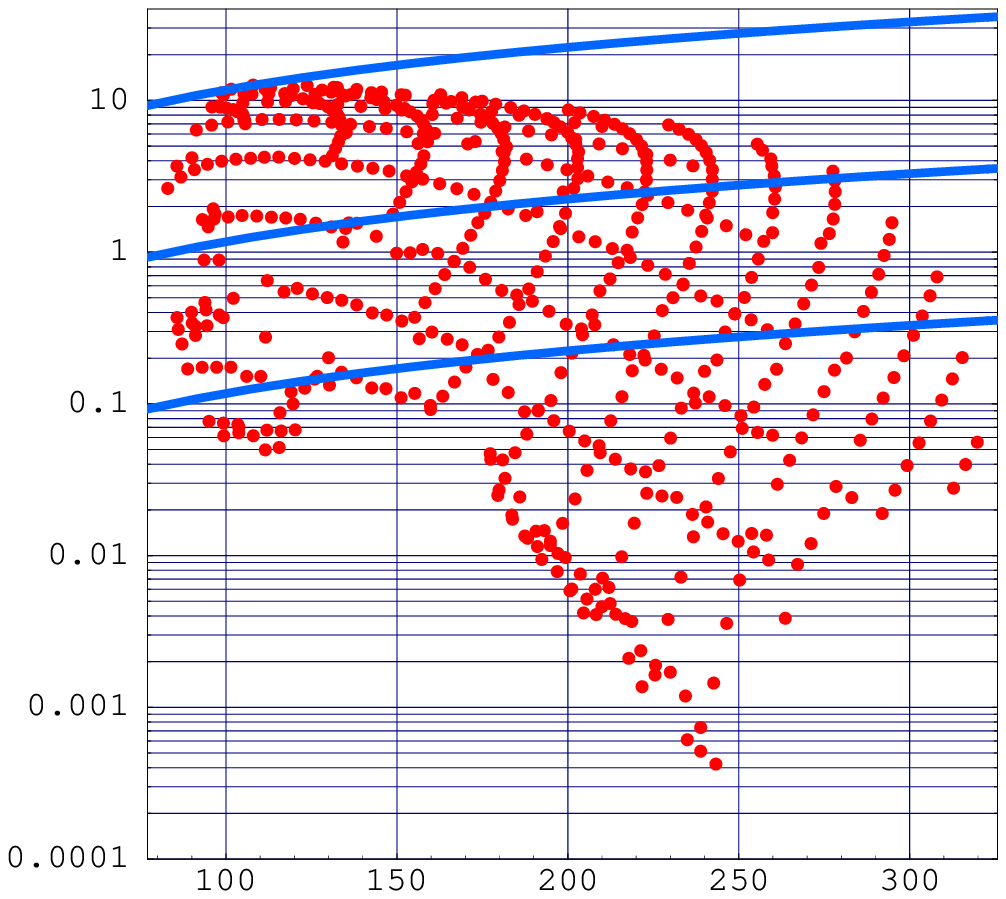,height=8cm}}
 %%%%%%%%%%%%%%%%%%%%%%%%%
 \begin{picture}(0,0)%%%%%
 %%%%%%%%%%%%%%%%%%%%%%%%%
  \put(55,150){$2 v\sigma_{2\gamma}$}
  \put(40,135){$[10^{-29}{\rm cm^3s^{-1}}]$}
  \put(353,160){${ j=10^{5}}$}
   \put(353,200){${ j=10^{4}}$}
   \put(353,240){${ j=10^{3}}$}
   \put(210,5){$m_{\chi}$[GeV]}
\normalsize
 %%%%%%%%%%%%%%%%%%%%%%%%%
 \end{picture}%%%%%%%%%%%%
 %%%%%%%%%%%%%%%%%%%%%%%%%
\vspace{0.2cm} 
\caption{Annihilation rate of neutralinos into the $2\gamma$ final
 state in the mSUGRA scenario for the red region  in
 Fig.~\ref{FIG-forcus15}, also with the chargino mass
bound.}
\label{FIG-indirect-forcus15}
%%%%%%%%%%%%%%%%%%%%%%%%%%%%%%%%%%%%%%%%%%%%%%%%%%%%%%%%%%%%
\end{figure}

Finally, we show the annihilation rate of neutralinos into the
$2\gamma$ final states in Fig.~\ref{FIG-indirect-forcus15}.  Here we
use the result of a full one-loop calculation presented in
Ref.~\cite{two-photon}.  The large $\widetilde{H}$ component of the
neutralino enhances the annihilation rate, which even reaches $2 v
\sigma_{2\gamma}\sim 10^{-28}\;{\rm cm^3\;sec^{-1}}$ in regions where
the LSP is almost pure $\widetilde{H}$.  The fact that the
$\widetilde{H}$-like neutralino gives such a large annihilation rate
was already known and investigated in
Refs.~\cite{two-photon,gamma-rays}.  However, in the previous works,
the thermal relic of neutralinos was assumed to provide an appropriate
mass density of dark matter. Therefore, the $\widetilde{H}$-like
neutralino only appears in the region with very large mass,
$m_{\chi}\gsim 500\GEV$, because of a large annihilation cross section.
In contrast, the late-time decays of Q-balls lead to relatively light
neutralino dark matter with a significant $\widetilde{H}$ component.
This is much preferable regarding naturalness as well as
detection.

The $\gamma$-ray flux in a given direction of the sky is obtained
by integrating the contributions along the line of sight (l.o.s.).
The result highly depends on models of the density profile of the halo.
Fortunately, all the model dependences can be factored out in terms of the
following dimensionless function:
\begin{equation}
J(\psi)=\frac{1}{8.5 \;{\rm kpc}}\cdot\left(
\frac{1}{0.3\GEV\;{\rm cm}^{-3}}
\right)^{2}\int_{\rm l.o.s.}\rho^2(l)dl(\psi)\;,
\label{J-parameter}
\end{equation}
where $\psi$ is the angle between the direction of the galactic centre
and that of observation; $\rho(l)$ denotes the density profile of the
halo along the l.o.s..
The $\gamma$-ray flux is written by this
function, the neutralino mass and the cross section:
\begin{equation}
\Phi_{\gamma}(\psi)\simeq  1.87\times 10^{-11}\left(
\frac{N_{\gamma}v\sigma}{10^{-29}\;{\rm cm}^{3}\;{\rm sec}^{-1}}\right)
\left(\frac{10\GEV}{m_{\chi}}\right)^{2}\cdot J(\psi)\;{\rm cm}^{-2}
\;{\rm sec}^{-1}\;{\rm sr}^{-1}\;,
\label{photon-flux}
\end{equation}
where $N_{\gamma}$ is the number of photons in the final state,
and $N_{\gamma}=2$ in the $\chi\chi\rightarrow \gamma\gamma$ annihilation.

Although the maximum flux will be obtained in the direction of the
galactic centre $\psi=0$, the experimentally relevant value
is the integral of $J(\psi)$ over the solid angle around $\psi=0$
determined by the
angular acceptance of a detector.  Therefore, the relevant function
we have to treat is
\begin{equation}
\left<J(0)\right>(\Delta \Omega)=\frac{1}{\Delta \Omega}
\int_{\Delta \Omega} J(\psi)\;d\Omega\;,
\label{Exp-J}
\end{equation}
where $\Delta \Omega$ is the angular acceptance of the detector.
The value of $\vev{J(0)}$ and its dependence on $\Delta\Omega$ in
various halo models are
discussed in Ref.~\cite{gamma-rays}.
A typical value is $\Delta \Omega =10^{-3}\;{\rm
sr}$ for a generic next-generation air Cherenkov telescope (ACT).
We define the value of $\vev{J(0)}$ averaged over this angular
acceptance as $j=\vev{J(0)}(10^{-3}\;{\rm sr})$.

In Fig.~\ref{FIG-indirect-forcus15}, we also plot $5\sigma$
sensitivity curves for typical next generation ACT
arrays~\cite{VERITAS,HESS,MAGIC} observing the
galactic centre, adopting several models of dark matter distribution,
including a Moore et al. profile~\cite{profile-moor} for $j=10^5$ and
Navarro, Frenk and White profile~\cite{profile-NFW} for $j=10^3$.  In
obtaining the sensitivities curves, we consider an instrument with a
$10^{9}\;{\rm cm}^2$ effective area and $15\%$ energy resolution,
assuming $10^{6}\;{\rm sec}$ exposure and using the standard estimation of
the background presented in Ref.~\cite{gamma-rays}.  We have also adopted a
hadronic rejection that is improved by a factor of $16$  with respect to the
current Whipple detector, following the arguments presented in this
reference.  One can see that by assuming somewhat cuspy dark matter
profiles of the halo, we have a relatively large possibility for
detecting monoenergetic $\gamma$-rays, especially for the region with
almost pure higgsino LSP.  

\begin{figure}[t!]%%%%%%%%%%%%%%%%%%%%%%%%%%%%%%%%%%%%%%%%%%%
%%%%%%%%%%%%%%%%%%%%%%%%%%%%%%%%%%%%%%%%%%%%%%%%%%%%%%%%%%%%
 \centerline{\psfig{figure=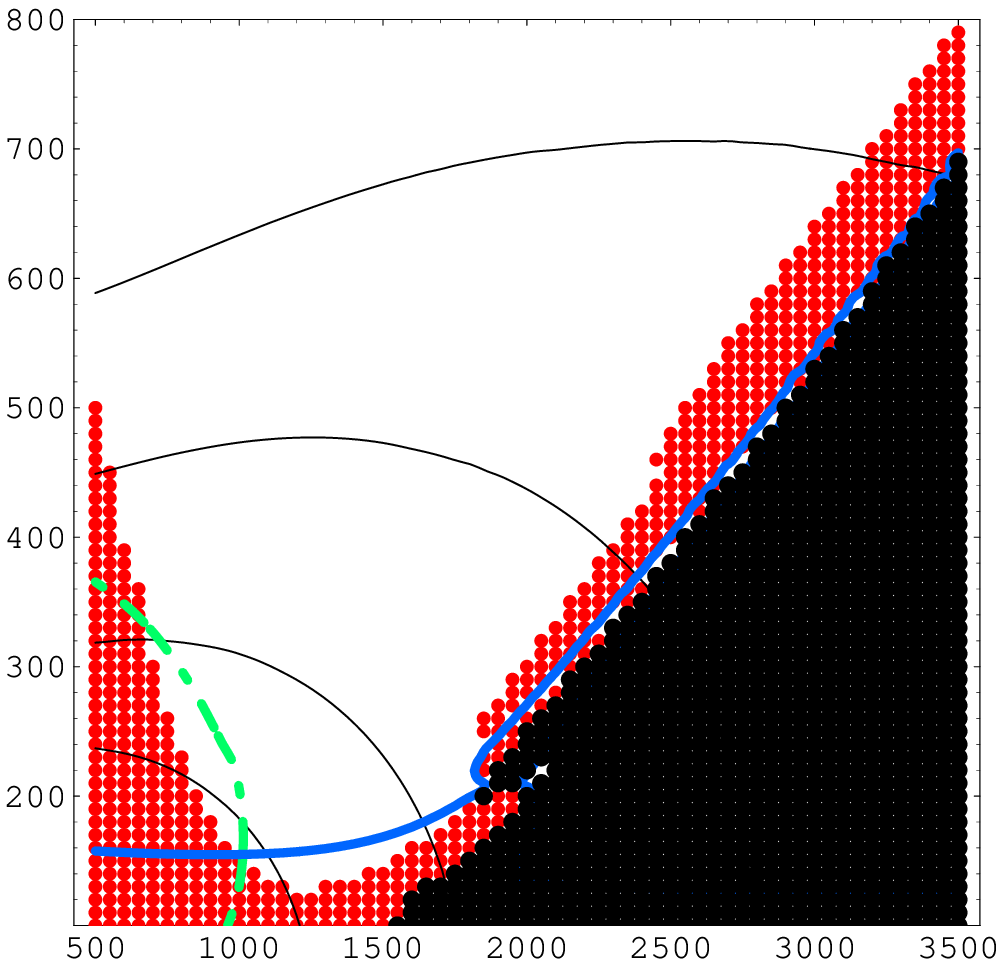,height=10cm}}
 %%%%%%%%%%%%%%%%%%%%%%%%%
 \begin{picture}(0,0)%%%%%
 %%%%%%%%%%%%%%%%%%%%%%%%
  \put(55,306){$M_{1/2}$[GeV]}  
   \put(200,1){$m_{0}$[GeV]}
\normalsize
 %%%%%%%%%%%%%%%%%%%%%%%%%
 \end{picture}%%%%%%%%%%%%
 %%%%%%%%%%%%%%%%%%%%%%%%%
\vspace{0.5cm}
 \caption{The allowed region in the mSUGRA scenario with ${\rm
tan}\beta=40$ and $A_{0}=0$ in $(m_{0}$--$M_{1/2})$ plane.  In the red shaded region, non-thermally produced LSPs
via decays of Q-balls result in a cosmologically interesting mass
density.  The black shaded region is where the electroweak symmetry
breaking cannot be implemented. The region below the blue (thick) line
is excluded by the chargino mass bound $m_{\chi^{\pm}}\gsim 105\GEV$.
The contours of the light Higgs boson mass are given by the black (thin)
lines, which correspond to $m_{h}=114.1, 117,\;120,\;122\GEV$,
respectively. The region below the green (dot-dashed) line is excluded 
where the branching ratio of $b\rightarrow s\gamma$ violates the CLEO bound,
$B(B\rightarrow X_{s}\gamma)>2\times 10^{-4}$.} \label{FIG-forcus40}
%%%%%%%%%%%%%%%%%%%%%%%%%%%%%%%%%%%%%%%%%%%%%%%%%%%%%%%%%%%%
\end{figure}

\begin{figure}[t!]%%%%%%%%%%%%%%%%%%%%%%%%%%%%%%%%%%%%%%%%%%%
%%%%%%%%%%%%%%%%%%%%%%%%%%%%%%%%%%%%%%%%%%%%%%%%%%%%%%%%%%%%
 \centerline{\psfig{figure=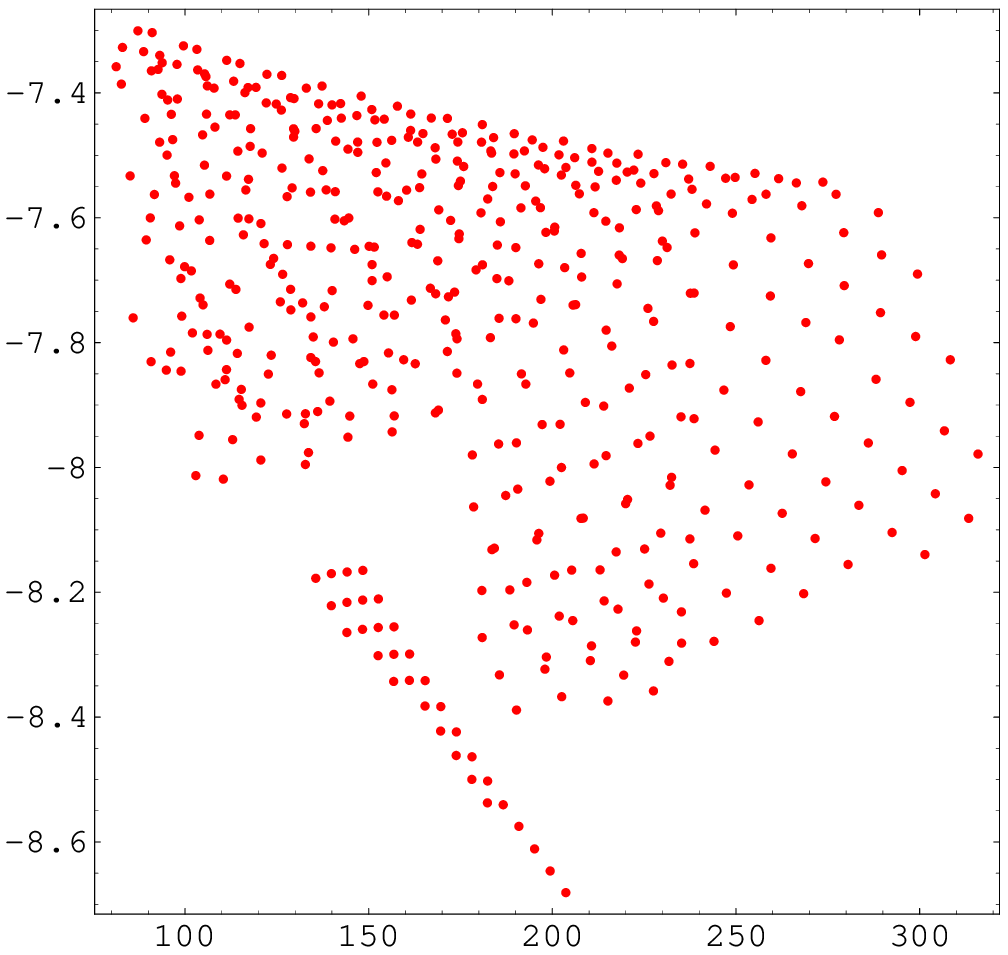,height=8cm}}
 %%%%%%%%%%%%%%%%%%%%%%%%%
 \begin{picture}(0,0)%%%%%
 %%%%%%%%%%%%%%%%%%%%%%%%%
  \put(43,140){${\rm log}_{10}\displaystyle{
\left(\frac{\sigma_{P}}{\rm [pb]}\right)}$}  
  \put(200,3){$m_{\chi}$[GeV]}
\normalsize
 %%%%%%%%%%%%%%%%%%%%%%%%%
 \end{picture}%%%%%%%%%%%%
 %%%%%%%%%%%%%%%%%%%%%%%%%
\vspace{0.cm}
 \caption{Proton--$\chi$ cross section in the  mSUGRA scenario for the 
red region  in
 Fig.~\ref{FIG-forcus40}, also with the chargino mass bound and the constraint from  $b\rightarrow s\gamma$.}
\label{FIG-direct-forcus40}
%%%%%%%%%%%%%%%%%%%%%%%%%%%%%%%%%%%%%%%%%%%%%%%%%%%%%%%%%%%%
\end{figure}

\begin{figure}[h!]%%%%%%%%%%%%%%%%%%%%%%%%%%%%%%%%%%%%%%%%%%%
%%%%%%%%%%%%%%%%%%%%%%%%%%%%%%%%%%%%%%%%%%%%%%%%%%%%%%%%%%%%
 \centerline{\psfig{figure=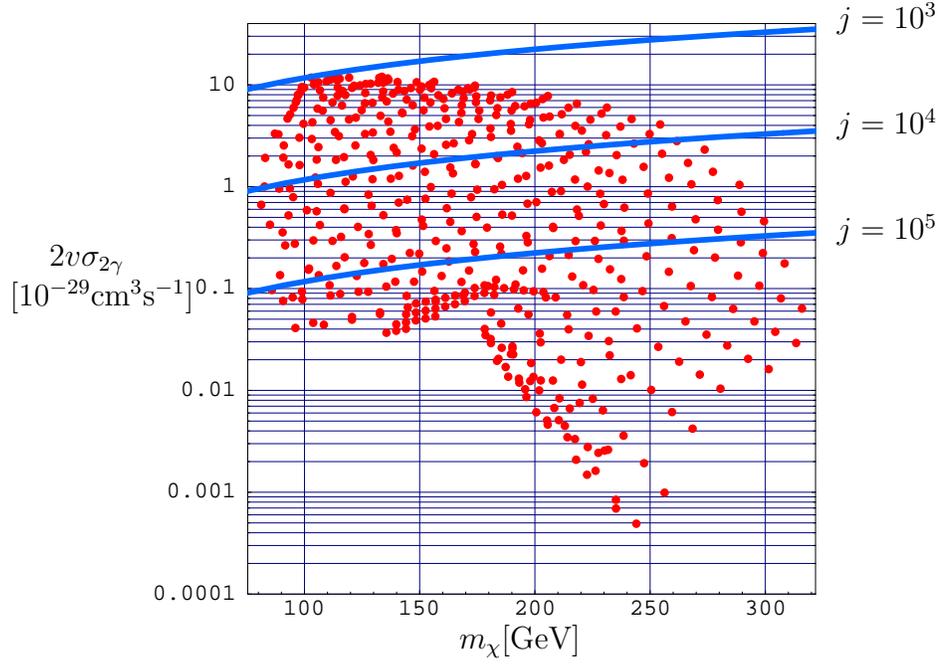,height=8cm}}
 %%%%%%%%%%%%%%%%%%%%%%%%%
 \begin{picture}(0,0)%%%%%
 %%%%%%%%%%%%%%%%%%%%%%%%%
  \put(55,150){$2 v\sigma_{2\gamma}$}
  \put(40,135){$[10^{-29}{\rm cm^3s^{-1}}]$}
  \put(353,160){${ j=10^{5}}$}
   \put(353,200){${ j=10^{4}}$}
   \put(353,240){${ j=10^{3}}$}
   \put(210,5){$m_{\chi}$[GeV]}
\normalsize
 %%%%%%%%%%%%%%%%%%%%%%%%%
 \end{picture}%%%%%%%%%%%%
 %%%%%%%%%%%%%%%%%%%%%%%%%
\vspace{0.2cm} 
\caption{Annihilation rate of neutralinos into the $2\gamma$ final state
 in the mSUGRA scenario for the red region  in
 Fig.~\ref{FIG-forcus40}, also with the chargino mass bound
 and the constraint from  $b\rightarrow s\gamma$.}  
\label{FIG-indirect-forcus40}
%%%%%%%%%%%%%%%%%%%%%%%%%%%%%%%%%%%%%%%%%%%%%%%%%%%%%%%%%%%%
\end{figure}

In Figs.~\ref{FIG-forcus40},~\ref{FIG-direct-forcus40} and
\ref{FIG-indirect-forcus40}, we show, respectively, the allowed
parameter space, the proton--$\chi$ cross sections and $\chi$-annihilation
rates into $2\gamma$ final state for the mSUGRA scenario with ${\rm
  tan}\beta=40$.  We take ${\rm sign}(\mu)$ to be positive also in
this case, which is desirable to avoid a large branching ratio of the
$b\rightarrow s\gamma$ transition.  Conventions are the same as in
Figs.~\ref{FIG-forcus15}, \ref{FIG-direct-forcus15} and
\ref{FIG-indirect-forcus15}. One can see some  differences in
Fig.~\ref{FIG-forcus40} compared with Fig.~\ref{FIG-forcus15}. First,
the region below the green (dot-dashed) line is excluded by the
constraints on the branching ratio of $b\rightarrow s\gamma$ from the CLEO
experiment, $B(B\rightarrow X_{s}\gamma)>2\times10^{-4}$.  Second, the
allowed region is significantly widened and even the region with the
bino-like LSP (the region for $m_{0}\lsim 1\TEV$) can lead to the
desired mass density of dark matter.  This is because, in the case of
large ${\rm tan}\beta$, the annihilation cross section through the 
process $\chi\chi\rightarrow A\rightarrow
f\bar{f}$ is strongly enhanced.  The reason is that the coupling with
fermions $Af\bar{f}$ is proportional to ${\rm tan}\beta$ and that the
mass of the pseudo-Higgs boson $m_{A}$ is significantly reduced by
large down-type Yukawa couplings, which results in a much smaller
suppression factor $(m_{A}/m_{\chi})^{4}$.

Here, we should add a comment.
One may think that the formation of Q-balls does not occur in
the focus point region, since a large soft mass squared $m_{0}^2$
would lead to the  $K$-factor
in Eq.~(\ref{potential-Q-ball-formation}) being positive.
However, even if $m_{0}^2={\cal O}(10\TEV)$, the $K$-factor is negative as
long as we adopt flat directions, which are solely constructed by
the first and second families. This condition can easily be  satisfied in most
of the flat directions.

\subsection{Parameter space and possibility of detection in
the anomaly-mediation model}
In pure anomaly-mediated SUSY breaking, soft terms are determined by
RG-invariant expressions involving the gauge and Yukawa couplings;
hence the soft terms are completely fixed by the low energy values
of these couplings and an overall scale $m_{3/2}$:
\begin{eqnarray}
&&M_{\lambda}=-\frac{g^2}{2}\frac{dg^{-2}}{d {\rm ln}\mu}m_{3/2}
=\frac{\beta_{g}}{g}m_{3/2}\;,\nonumber\\
&&m_{\tilde{\Phi}}^{2}=-\frac{1}{4}
\frac{d^2 {\rm ln}Z_{\Phi}}{d ({\rm ln\mu})^2}m_{3/2}^2=-\frac{1}{4}
\left(\frac{\partial \gamma}{\partial g}\beta_{g}+
\frac{\partial \gamma}{\partial y}\beta_{y}\right)m_{3/2}^2\;,
\nonumber\\
&&A_{y}=\frac{1}{2}\sum_{i}\frac{d{\rm ln} Z_{\Phi_{i}}}{d{\rm ln}y}m_{3/2}=
-\frac{\beta_{y}}{y}m_{3/2}\;,
\label{pure-anomaly}
\end{eqnarray}
where $g,\;y$ are gauge and Yukawa couplings, respectively,
$\Phi$ denotes the general superfield, and $m_{\Phi}^2,\;Z_{\Phi}$
are its soft scalar mass squared and wave function renormalization factor;
$M_{\lambda}$ is the soft mass for the gaugino $\lambda$,
$A_{y}$ is the trilinear scalar coupling associated with the Yukawa
interaction with a coupling constant $y$. 
Here the sum $\sum_{i}$ runs through the fields included in 
this Yukawa interaction.
We have defined the renormalization group functions as
$\gamma(g,y)\equiv d{\rm ln}Z/d {\rm ln}\mu$, $\beta_{g}(g,y)\equiv
dg/d{\rm ln}\mu$, and $\beta_{y}\equiv dy/d{\rm ln}\mu$.

As you can see from these relations, the soft masses squared for sleptons
are negative, which is the biggest difficulty in pure AMSB models.
Although many possible solutions are proposed~\cite{AMSB-cure}, we adopt,
in the present work, a
phenomenological solution to the negative slepton mass problem.
We assume the additional universal scalar mass $m_{0}$ at the GUT scale,
\begin{equation}
m_{\Phi}^2=-\frac{1}{4}
\left(\frac{\partial \gamma}{\partial g}\beta_{g}+
\frac{\partial \gamma}{\partial y}\beta_{y}\right)m_{3/2}^2+m_{0}^2\;,
\label{boundary}
\end{equation}
and then evolve the RG equations to obtain the low energy spectrum.
Then the entire parameter space is specified by the following $4$
parameters in this minimal framework:
\begin{equation}
m_{3/2},\;m_{0},\;{\rm tan}\beta,\;{\rm sign}(\mu)\;.
\label{AMSB-parameters}
\end{equation}

{}From Eq.~(\ref{pure-anomaly}), we see that the ratios between the
gaugino masses at the weak scale are approximately given by
\begin{equation}
M_{1}\;:\;M_{2}\;:\;M_{3}\approx 2.8\;:\;1\;:\;-8.3\;,
\end{equation}
where $M_{1},\;M_{2}$ and $M_{3}$ are the gaugino masses
for U(1)$_{Y}$, SU(2)$_{L}$ and SU(3)$_{C}$, respectively.
Therefore, in the present scenario, the $\widetilde{W}$ LSP is realized
in most of the parameter space.

\begin{figure}[t!]%%%%%%%%%%%%%%%%%%%%%%%%%%%%%%%%%%%%%%%%%%%
%%%%%%%%%%%%%%%%%%%%%%%%%%%%%%%%%%%%%%%%%%%%%%%%%%%%%%%%%%%%
 \centerline{\psfig{figure=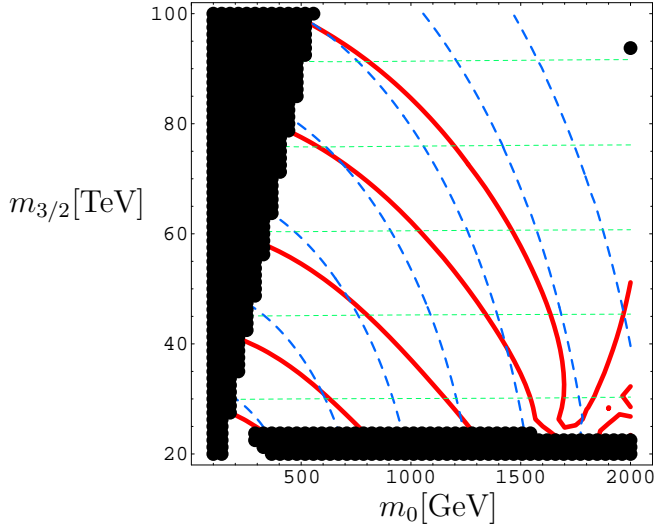,height=6.5cm}}
 %%%%%%%%%%%%%%%%%%%%%%%%%
 \begin{picture}(0,0)%%%%%
 %%%%%%%%%%%%%%%%%%%%%%%%%
  \put(70,120){$m_{3/2}$[TeV]}
  \put(210,5){$m_{0}$[GeV]}
\normalsize
 %%%%%%%%%%%%%%%%%%%%%%%%%
 \end{picture}%%%%%%%%%%%%
 %%%%%%%%%%%%%%%%%%%%%%%%%
\vspace{0.2cm} 
\caption{Allowed parameter space and proton--$\chi$ cross section in 
the anomaly-mediation model with ${\rm tan\beta}=15$.  
Here, we take ${\rm sign}(\mu)$ negative.
In this plot, there is no region excluded by $b\rightarrow s\gamma$.
The black shaded region is
excluded by $\tilde{\tau}$-LSP or the produced neutralinos overclose the
universe.  The wide white region leads to a desired mass density of dark
matter via decays of Q-balls.  The red (thick)
lines are contours of proton--$\chi$ cross section, which are
$\sigma_{P}=10^{-8}$, $10^{-9}$, $10^{-10}$, $10^{-11}$ and
$10^{-12}$ pb,  from left to right, respectively.  The blue (dashed)
lines are contours of the 
mass of the heavy Higgs boson $m_{H}$, which are,
from left to right, $500,\;750,\;1000,\;\ldots\;2000\GEV$.  The green
(dotted) lines are contours of the lightest neutralino mass $m_{\chi}$
for $100,\;150,\ldots\;300\GEV$.}
\label{FIG-anomaly15}
%%%%%%%%%%%%%%%%%%%%%%%%%%%%%%%%%%%%%%%%%%%%%%%%%%%%%%%%%%%%
\end{figure}

In Fig.~\ref{FIG-anomaly15}, we present a parameter space where the LSPs
produced by late-time decays of Q-balls provide a desired mass density
of dark matter and the proton--$\chi$ scalar cross section in that region.
Here we take ${\rm tan}\beta=15$, and ${\rm sign}(\mu)$ negative to
avoid too large a branching ratio of $b\rightarrow s\gamma$.  There is
no region excluded from the constraints on $b\rightarrow s\gamma$.  The
black shaded region is excluded by the $\tilde{\tau}$ LSP and/or
overclosing universe by the LSPs produced via Q-ball decays. In all of
the remaining white space, the $\widetilde{W}$ LSP from the Q-ball decay
can be the dominant component of the dark matter. The typical decay
temperature of Q-balls resulting in the suitable mass density of dark
matter is $10\MEV\lsim T_{d}\lsim 100\MEV$. On the other hand, the mass
splitting between the charged and neutral $\widetilde{W}$'s is of order
$100\MEV$ to $1\GEV$~\cite{Ghel-Giudice,Feng-Moroi-Su}, which allows us to
neglect the co-annihilation effects.

We present contours of the proton--$\chi$
cross section by the red (thick) lines, which correspond to
$\sigma_{P}=10^{-8},\;10^{-9},\ldots,\; 10^{-12}$ pb from left to
right, respectively. A strange structure appearing in the lower 
right-hand corner indicates a contamination of the $\widetilde{H}$ component
in the LSP.  The green (dotted) lines are the contours of the lightest
neutralino mass, $m_{\chi}=100,\;150,\ldots,\;300\GEV$, respectively.
We also present the contours of heavy Higgs boson mass by the blue
(dashed) lines, which denote $m_{H}=500,\;750,\ldots,\;2000\GEV$ from
left to right, respectively.  In contrast to the previous
expectation~\cite{FH,Moroi-Randall}, the direct detection rates are
rather small, and only a restricted region is within the reach of the
next-generation detectors, $\sigma_{P}\gsim 10^{-(9-8)}$ pb. This
comes from the large mass of the heavy Higgs boson $m_{H}$. In
relatively large ${\rm tan\beta}$, the proton--$\chi$ cross section is
dominated by $H$ exchange, and it reduces as $\sigma_{P}\propto
m_{H}^{-4}$.  In Refs.~\cite{FH,Moroi-Randall}, it was assumed that
$m_{H}\simeq 300\GEV$, which is realized only in a small space, as one
can see from Fig.~\ref{FIG-anomaly15}.

\begin{figure}[h!]%%%%%%%%%%%%%%%%%%%%%%%%%%%%%%%%%%%%%%%%%%%
%%%%%%%%%%%%%%%%%%%%%%%%%%%%%%%%%%%%%%%%%%%%%%%%%%%%%%%%%%%%
 \centerline{\psfig{figure=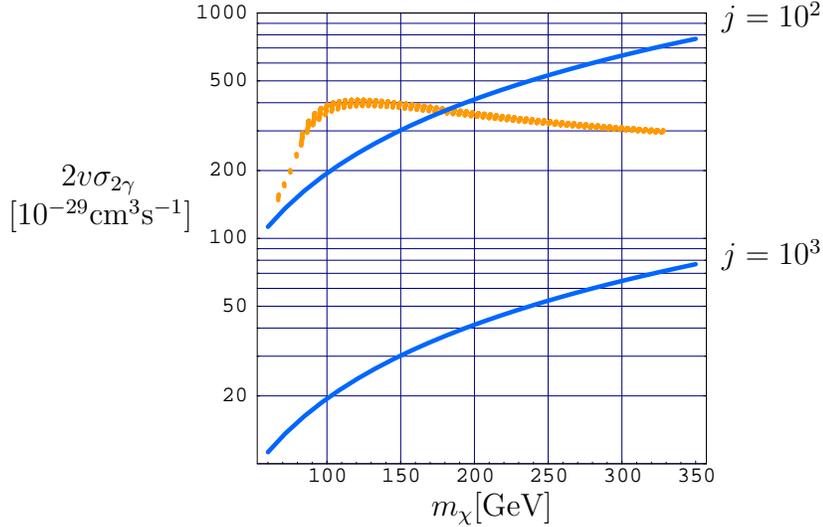,height=6.5cm}}
 %%%%%%%%%%%%%%%%%%%%%%%%%
 \begin{picture}(0,0)%%%%%
 %%%%%%%%%%%%%%%%%%%%%%%%%
    \put(70,130){$2 v\sigma_{2\gamma}$}
  \put(50,115){$[10^{-29}{\rm cm^3s^{-1}}]$}
  \put(320,100){$j=10^{3}$}
  \put(320,190){$j=10^{2}$}
   \put(210,5){$m_{\chi}$[GeV]}
\normalsize
 %%%%%%%%%%%%%%%%%%%%%%%%%
 \end{picture}%%%%%%%%%%%%
 %%%%%%%%%%%%%%%%%%%%%%%%%
\vspace{0.2cm} 
\caption{Annihilation rate of neutralinos into the $2\gamma$ final state
 in the AMSB model with ${\rm tan}\beta=15$.
Each orange dot corresponds to one parameter set in the white region
in Fig.~\ref{FIG-anomaly15}. The two blue lines denote the $5\sigma$
 sensitivity curves for $j=10^2,\;10^3$.
}
\label{FIG-indirect-anomaly15}
%%%%%%%%%%%%%%%%%%%%%%%%%%%%%%%%%%%%%%%%%%%%%%%%%%%%%%%%%%%%
\end{figure}

Fortunately, we may obtain a clear signal from the
neutralino annihilation into the $2\gamma$ final state~\cite{AMSB-photon}.
In Fig.~\ref{FIG-indirect-anomaly15}, we show the annihilation rate of
the LSP into the $2\gamma$ final state. Each orange dot falls in
somewhere in the allowed space presented in Fig.~\ref{FIG-anomaly15}.
The blue lines are $5\sigma$ sensitivity curves for $j=10^2$ and $10^3$,
which are estimated by the same parameter set as was used in
Figs.~\ref{FIG-indirect-forcus15} and \ref{FIG-indirect-forcus40},
but we do not assume any improvements on the hadronic rejection
and use $\epsilon_{\rm had}=1$ in Ref.~\cite{gamma-rays}.
Even if we do not assume strongly cuspy profiles of the dark matter
density in the halo, a large portion of the parameter space predicts
a measurable size of the $\gamma$-ray flux.

\begin{figure}[t!]%%%%%%%%%%%%%%%%%%%%%%%%%%%%%%%%%%%%%%%%%%%
%%%%%%%%%%%%%%%%%%%%%%%%%%%%%%%%%%%%%%%%%%%%%%%%%%%%%%%%%%%%
 \centerline{\psfig{figure=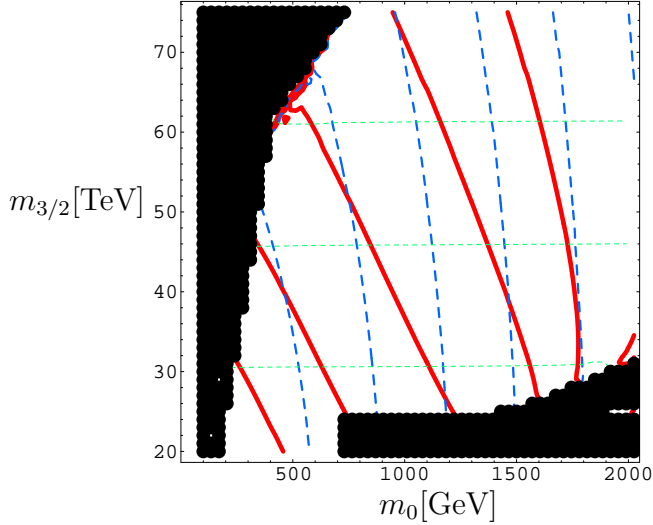,height=6.5cm}}
 %%%%%%%%%%%%%%%%%%%%%%%%%
 \begin{picture}(0,0)%%%%%
 %%%%%%%%%%%%%%%%%%%%%%%%%
  \put(70,120){$m_{3/2}$[TeV]}
  \put(210,5){$m_{0}$[GeV]}
\normalsize
 %%%%%%%%%%%%%%%%%%%%%%%%%
 \end{picture}%%%%%%%%%%%%
 %%%%%%%%%%%%%%%%%%%%%%%%%
\vspace{0.2cm} 
\caption{Allowed parameter space and proton--$\chi$ cross section in 
the anomaly-mediation model with ${\rm tan\beta}=30$.  
Here, we take ${\rm sign}(\mu)$ negative.
The black shaded region is
excluded by $\tilde{\tau}$-LSP or the produced neutralinos overclose the
universe.  The black region also includes small regions where we cannot
obtain the convergence of SOFTSUSY code.
The wide white region leads to a desired mass density of dark
matter via decays of Q-balls.  The red (thick)
lines are contours of proton--$\chi$ cross section, which are
$\sigma_{P}=10^{-7}$, $10^{-8}$, $10^{-9}$, $10^{-10}$ and
$10^{-11}$ pb, from left to right, respectively.  The blue (dashed)
lines are contours of the mass of the heavy Higgs boson $m_{H}$, which are,
from left to right, $500,\;750,\;1000,\;\ldots\;1750\GEV$.  The green
(dotted) lines are contours of the lightest neutralino mass $m_{\chi}$
for $100,\;150,\;200\GEV$.}
\label{FIG-anomaly30}
%%%%%%%%%%%%%%%%%%%%%%%%%%%%%%%%%%%%%%%%%%%%%%%%%%%%%%%%%%%%
\end{figure}

\begin{figure}[h!]%%%%%%%%%%%%%%%%%%%%%%%%%%%%%%%%%%%%%%%%%%%
%%%%%%%%%%%%%%%%%%%%%%%%%%%%%%%%%%%%%%%%%%%%%%%%%%%%%%%%%%%%
 \centerline{\psfig{figure=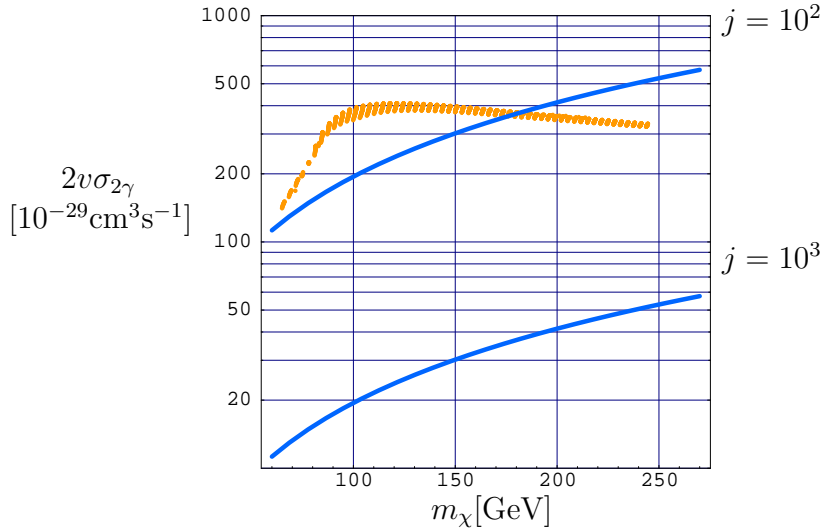,height=6.5cm}}
 %%%%%%%%%%%%%%%%%%%%%%%%%
 \begin{picture}(0,0)%%%%%
 %%%%%%%%%%%%%%%%%%%%%%%%%
    \put(70,130){$2 v\sigma_{2\gamma}$}
  \put(50,115){$[10^{-29}{\rm cm^3s^{-1}}]$}
  \put(320,100){$j=10^{3}$}
  \put(320,190){$j=10^{2}$}
   \put(210,5){$m_{\chi}$[GeV]}
\normalsize
 %%%%%%%%%%%%%%%%%%%%%%%%%
 \end{picture}%%%%%%%%%%%%
 %%%%%%%%%%%%%%%%%%%%%%%%%
\vspace{0.2cm} 
\caption{Annihilation rate of neutralinos into the $2\gamma$ final state
 in the AMSB model with ${\rm tan}\beta=30$.
Conventions are the same as those in Fig~\ref{FIG-indirect-anomaly15}.
}
\label{FIG-indirect-anomaly30}
%%%%%%%%%%%%%%%%%%%%%%%%%%%%%%%%%%%%%%%%%%%%%%%%%%%%%%%%%%%%
\end{figure}

In Figs.~\ref{FIG-anomaly30} and \ref{FIG-indirect-anomaly30}, we show
the plots for the case of ${\rm tan}\beta=30$ and ${\rm sign}(\mu)<0$.
Conventions are almost the same as those in Figs.~\ref{FIG-anomaly15}
and \ref{FIG-indirect-anomaly15}, respectively, but they are explicitly
denoted in the caption to each figure.  In such a large ${\rm
  tan}\beta$, we can see that the proton--$\chi$ cross section is
strongly enhanced.  This is because the $H$--nucleon--nucleon
coupling, which dominates the proton--$\chi$ cross section, is
proportional to ${\rm tan}\beta$, and the large down-type Yukawa
couplings reduce the mass of the heavy Higgs boson $m_{H}^2$.  We can
expect the direct detection of the $\widetilde{W}$ dark matter in
a relatively wide parameter space in the next generation of detectors.  In
Fig.~\ref{FIG-anomaly30}, a small region with $m_{0}\lsim
500\GEV,\;m_{3/2}\lsim 30\TEV$ is excluded by the constraint
$B(B\rightarrow X_{s}\gamma)>2\times 10^{-4}$~\cite{feng-moroi-AMSB},
although it is not depicted in the figure.  The neutralino
annihilation rate into the $2\gamma$ final state is almost the same as
in the case of ${\rm tan\beta}=15$, which is consistent with the
result of Ref.~\cite{AMSB-photon}.

In AMSB models, the neutralino annihilation into the $Z\gamma$ final
state is known to be nearly twice that into the $2\gamma$ final
state (see Ref.~\cite{AMSB-photon}). Note that, in such a minimal framework of
AMSB models, the thermal relic of the $\widetilde{W}$ cannot be a
significant component of dark matter, because of its large annihilation
cross section. Such a large annihilation cross section of the LSP is,
on the contrary, much more advantageous for the AD baryogenesis with
late-time decays of Q-balls, although the conditions given in
Eq.~(\ref{ADinAMSBwithsuper}) or (\ref{ADinAMSBwithoutsuper}) should
be satisfied.

\subsection{Parameter space and possibility of detection in
no-scale models\\ with non-universal gaugino masses}

In no-scale models~\cite{no-scale-org}, all the soft parameters except gaugino masses are
assumed to vanish at some high energy scale $M_{X}$, which is usually
taken to be the GUT scale $M_{X}=M_{G}$.  The soft parameters at the
weak scale are generated by RG effects dominated by the gaugino masses,
which are automatically flavour-blind and naturally suppress the SUSY
flavour-changing neutral current (FCNC) interactions~\cite{kawasaki-yanagida}.
Recently, models with the no-scale boundary condition have attracted much
attention, since a natural realization was found as gaugino-mediation
models~\cite{gaugino-MSB}.

Unfortunately, the MSSM with the no-scale boundary condition at the
GUT scale was found to be inconsistent with phenomenological
requirements~\cite{Komine-Yamaguchi,Komine,EllisLower}.  This is mainly due to the lower bound on the light Higgs
boson mass and the cosmological requirement that a charged particle
must not be the LSP. Several solutions were proposed to  this
problem, which are, imposing the no-scale boundary conditions above
the GUT scale~\cite{above-GUT}, gauging the U(1)$_{B-L}$
symmetry~\cite{Fujii-Suzuki} and assuming the non-universal gaugino
masses at the GUT scale~\cite{Komine-Yamaguchi}.

In the present work, we pick up the last scenario with non-universal
gaugino masses. In this scenario, as investigated in
Ref.~\cite{Komine-Yamaguchi}, the $\widetilde{W}$ LSP is realized in a
wide parameter space, which is expected to derive a desired mass
density of dark matter from late-time decays of Q-balls.

\begin{figure}[t!]%%%%%%%%%%%%%%%%%%%%%%%%%%%%%%%%%%%%%%%%%%%
%%%%%%%%%%%%%%%%%%%%%%%%%%%%%%%%%%%%%%%%%%%%%%%%%%%%%%%%%%%%
 \centerline{\psfig{figure=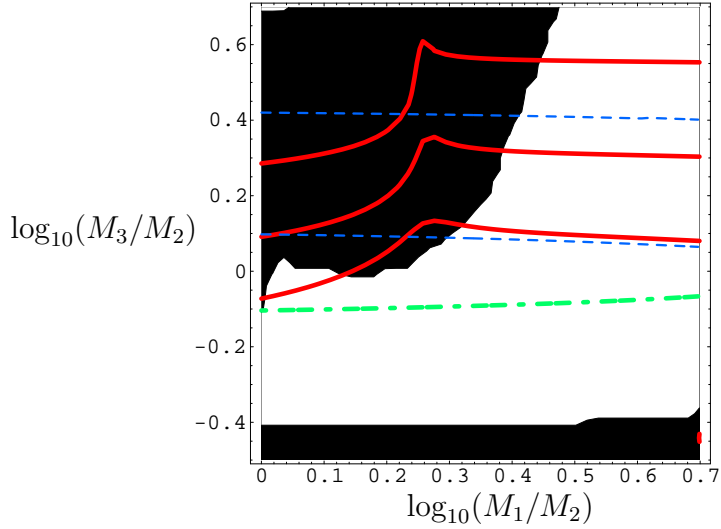,height=6.5cm}}
 %%%%%%%%%%%%%%%%%%%%%%%%%
 \begin{picture}(0,0)%%%%%
 %%%%%%%%%%%%%%%%%%%%%%%%%
  \put(50,110){${\rm log}_{10}(M_{3}/M_{2})$}
  \put(200,5){${\rm log}_{10}(M_{1}/M_{2})$}
\normalsize
 %%%%%%%%%%%%%%%%%%%%%%%%%
 \end{picture}%%%%%%%%%%%%
 %%%%%%%%%%%%%%%%%%%%%%%%%
\vspace{0.2cm} 
\caption{Allowed region and the proton--$\chi$ scalar cross section
 $\sigma_{P}$ in the no-scale model with non-universal gaugino masses with
 ${\rm tan}\beta=10$.  
Here, we take $M_{2}=200\GEV$ at the GUT scale.
The red (thick) lines are the contours of the
 $\sigma_{P}=10^{-7}$, $10^{-8}$, $10^{-9}$ pb from the bottom up,
 respectively. The black shaded region is excluded by the fact that 
 $\tilde{\tau}$ is the LSP, or the resultant LSPs from late-time
 decays of Q-balls overclose the universe, or the EWSB cannot be 
implemented.  The blue (dashed) lines are
 the contours of the Higgs boson mass $m_{h} =114.1,\;120\GEV$ and the
 region below the lower line is excluded. The region below the green (dot-dashed) line is excluded since
 $B(B\rightarrow X_{s}\gamma)<2 \times 10^{-4}$.  } 
 \label{FIG-noscale10}
%%%%%%%%%%%%%%%%%%%%%%%%%%%%%%%%%%%%%%%%%%%%%%%%%%%%%%%%%%%%
\end{figure}

In Fig.~\ref{FIG-noscale10}, we present allowed regions where the
resultant LSPs give a desired mass density of dark matter and the
proton--$\chi$ cross section in that region. Here, we take
$M_{2}=200\GEV$, ${\rm tan}\beta=10$ 
and vary the ratios of the gaugino masses at the GUT
scale $M_{1}/M_{2}$ and $M_{3}/M_{2}$, which are denoted by the $x$-axis
and the $y$-axis, respectively.  The red (thick) solid lines are the
contours of the proton--$\chi$ cross section $\sigma_{P}$, which are
$\sigma_{P}=10^{-7}$, $10^{-8}$ and $10^{-9}$ pb
from the bottom up. The black shaded region is excluded,
since the $\tilde{\tau}$ is the LSP, or the resultant LSPs
overclose the universe, or the EWSB cannot be implemented. The
blue (dashed) lines are the contours of the Higgs boson mass,
$m_{h}=114.1$ and $120\GEV$, and the region below the lower line is
excluded by the Higgs boson mass bound~\cite{higgs-search}.  The region below the green
(dot-dashed) line is excluded by the constraints on the $b\rightarrow
s\gamma$, $B(B\rightarrow X_{s}\gamma)>2\times 10^{-4}$.

\begin{figure}[h!]%%%%%%%%%%%%%%%%%%%%%%%%%%%%%%%%%%%%%%%%%%%
%%%%%%%%%%%%%%%%%%%%%%%%%%%%%%%%%%%%%%%%%%%%%%%%%%%%%%%%%%%%
 \centerline{\psfig{figure=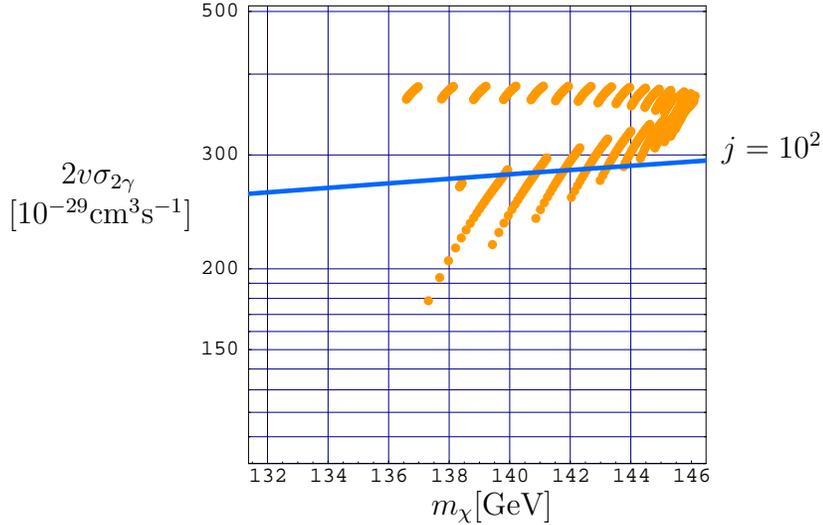,height=6.5cm}}
 %%%%%%%%%%%%%%%%%%%%%%%%%
 \begin{picture}(0,0)%%%%%
 %%%%%%%%%%%%%%%%%%%%%%%%%
    \put(70,130){$2 v\sigma_{2\gamma}$}
  \put(50,115){$[10^{-29}{\rm cm^3s^{-1}}]$}
  \put(320,140){$j=10^{2}$}
   \put(210,5){$m_{\chi}$[GeV]}
\normalsize
 %%%%%%%%%%%%%%%%%%%%%%%%%
 \end{picture}%%%%%%%%%%%%
 %%%%%%%%%%%%%%%%%%%%%%%%%
\vspace{0.2cm} 
\caption{Annihilation rate of neutralinos into the $2\gamma$ final state
 for the allowed region in Fig.~\ref{FIG-noscale10}. Conventions 
are the same as those in Fig.~\ref{FIG-indirect-anomaly15}.
}
\label{FIG-indirect-noscale10}
%%%%%%%%%%%%%%%%%%%%%%%%%%%%%%%%%%%%%%%%%%%%%%%%%%%%%%%%%%%%
\end{figure}

In Fig.~\ref{FIG-indirect-noscale10}, we show the neutralino
annihilation cross section into the $2\gamma$ final state for the
allowed region in Fig.~\ref{FIG-noscale10}. We also present the
$5\sigma$ sensitivity curve for $j=10^2$. Conventions are the same as
those in Fig.~\ref{FIG-indirect-anomaly15} for anomaly-mediation
models.  In the allowed region, the LSP is given by the
$\widetilde{W}$-like neutralino, which predicts large annihilation
cross sections comparable with those in anomaly-mediation models.

\begin{figure}[t!]%%%%%%%%%%%%%%%%%%%%%%%%%%%%%%%%%%%%%%%%%%%
%%%%%%%%%%%%%%%%%%%%%%%%%%%%%%%%%%%%%%%%%%%%%%%%%%%%%%%%%%%%
 \centerline{\psfig{figure=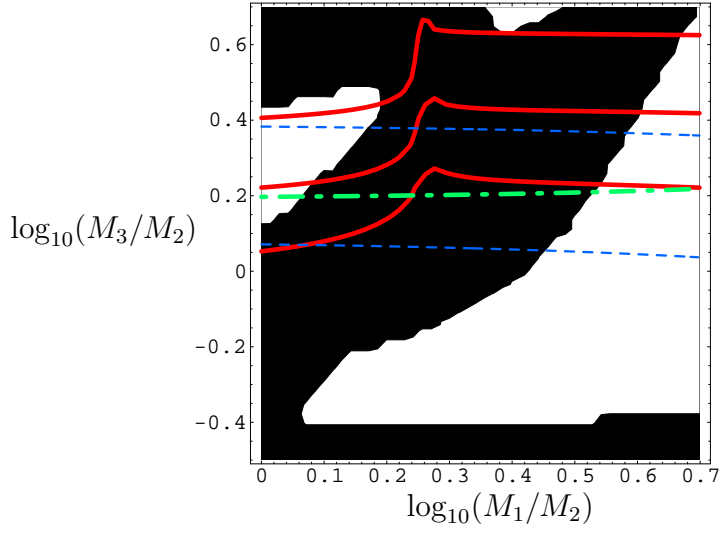,height=6.5cm}}
 %%%%%%%%%%%%%%%%%%%%%%%%%
 \begin{picture}(0,0)%%%%%
 %%%%%%%%%%%%%%%%%%%%%%%%%
 \put(50,110){${\rm log}_{10}(M_{3}/M_{2})$}
  \put(200,5){${\rm log}_{10}(M_{1}/M_{2})$}
\normalsize
 %%%%%%%%%%%%%%%%%%%%%%%%%
 \end{picture}%%%%%%%%%%%%
 %%%%%%%%%%%%%%%%%%%%%%%%%
\vspace{0.2cm} 
\caption{Allowed region and the proton--$\chi$ scalar cross section
 $\sigma_{P}$ in the no-scale model with non-universal gaugino masses with
 ${\rm tan}\beta=30$. Conventions are the same as those in 
Fig.~\ref{FIG-noscale10}. The bino-like LSP is realized in the small
 spot appearing on the left of
 the excluded region. The required large annihilation cross section is 
obtained by the $A$ exchange diagram enhanced by the 
large ${\rm tan}\beta$ and the  relatively small $m_{A}$. 
}
\label{FIG-noscale30}
%%%%%%%%%%%%%%%%%%%%%%%%%%%%%%%%%%%%%%%%%%%%%%%%%%%%%%%%%%%%
\end{figure}

\begin{figure}[h!]%%%%%%%%%%%%%%%%%%%%%%%%%%%%%%%%%%%%%%%%%%%
%%%%%%%%%%%%%%%%%%%%%%%%%%%%%%%%%%%%%%%%%%%%%%%%%%%%%%%%%%%%
 \centerline{\psfig{figure=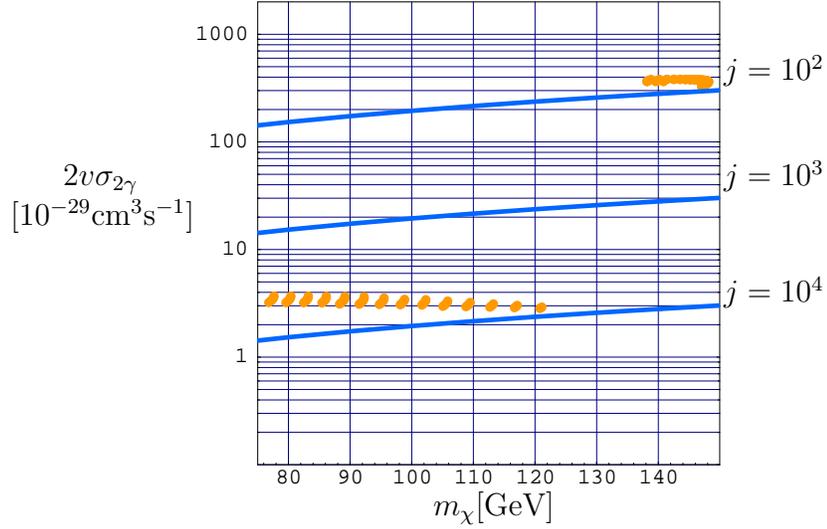,height=6.5cm}}
 %%%%%%%%%%%%%%%%%%%%%%%%%
 \begin{picture}(0,0)%%%%%
 %%%%%%%%%%%%%%%%%%%%%%%%%
    \put(70,130){$2 v\sigma_{2\gamma}$}
  \put(50,115){$[10^{-29}{\rm cm^3s^{-1}}]$}
  \put(320,86){$j=10^{4}$}
  \put(320,130){$j=10^{3}$}
  \put(320,170){$j=10^{2}$}
   \put(210,5){$m_{\chi}$[GeV]}
\normalsize
 %%%%%%%%%%%%%%%%%%%%%%%%%
 \end{picture}%%%%%%%%%%%%
 %%%%%%%%%%%%%%%%%%%%%%%%%
\vspace{0.2cm} 
\caption{Annihilation rate of neutralinos into the $2\gamma$ final state
for the allowed region in Fig.~\ref{FIG-noscale30}.
Conventions are the same as those in Fig~\ref{FIG-indirect-anomaly15}.
The dots with smaller annihilation rates correspond to the region with  the
bino-like LSP.
}
\label{FIG-indirect-noscale30}
%%%%%%%%%%%%%%%%%%%%%%%%%%%%%%%%%%%%%%%%%%%%%%%%%%%%%%%%%%%%
\end{figure}

In Figs.~\ref{FIG-noscale30} and \ref{FIG-indirect-noscale30}, we
present the corresponding figures in the case of ${\rm tan}\beta=30$.
Conventions are the same as those in Figs.~\ref{FIG-noscale10} and
\ref{FIG-indirect-noscale10}. ($M_{2}=200\GEV$ is assumed at the GUT
scale.)  In this case, the direct detection rate is enhanced by large
${\rm tan}\beta$, but a large portion of the parameter space is excluded
by the constraint from the $b\rightarrow s\gamma$.  The bino-like LSP is
realized in the small spot appearing on the left of the excluded region.
In this region, the large annihilation cross section required from the
late-time decays of Q-balls is provided by the pseudo-Higgs $A$ exchange
diagram via the small $\widetilde{H}$ contamination in the LSP,
whose amplitude is enhanced by the large ${\rm tan}\beta=30$
and the relatively small $m_{A}$.
In this bino-like LSP region, the annihilation rate into the $2\gamma$
final state is relatively small $2v\sigma_{2\gamma}\sim 3\times
10^{-29}\;{\rm cm}^{3}\;{\rm sec}^{-1}$.

\subsection*{Warm dark matter production from late-time decays of Q-balls}

Before closing this section, we should remark on an
interesting possibility that the resultant
$\widetilde{H}$-like LSPs  (or maybe $\widetilde{B}$-like LSP in the
case of a large ${\rm tan}\beta$) form the warm dark matter (WDM).
As we have seen, cuspy structures in the halo profiles have
a drastic consequence on the future observations in indirect
dark matter searches. However, it is argued that the neutralino
cuspy profile might be inconsistent with the radio emission from
the centre of the Galaxy~\cite{cuspy?}, although further detailed
studies are needed to fix the situation. Maybe the inflation models
with a suppressed strength in the small-scale perturbations might solve the
problem.
Another possible solution is the WDM.

If the neutralino LSPs are produced after the decoupling of thermal
interactions with relativistic velocity, they may serve as a WDM and
wash out the cuspy profiles in the
halo~\cite{Non-thermal-WIMP0,Non-thermal-WIMP}.  The conditions needed
for the MSSM neutralino to form a WDM were studied in
Ref.~\cite{neutralino-WDM}.  Actually, these authors found that it is
possible if the temperature is low enough when the neutralinos are
produced via decays of non-thermal sources:
\begin{eqnarray}
&&T_{d}\lsim 5\MEV\;\; {\mbox{for}}\;\; \widetilde{B}{\mbox{-like LSP}}
\nonumber\\
{\mbox{and}}\quad &&\nonumber\\
&&T_{d}\lsim 2\MEV \;\;{\mbox{for}}\;\; \widetilde{H}{\mbox{-like LSP}}.
\label{wdm-0}
\end{eqnarray}
As for the $\widetilde{W}$ LSP, they found it is impossible
to form the WDM.
If the above condition is satisfied, energy reductions of neutralinos
by scatterings with thermal background are small enough.
Such a situation may arise in some models,
for example,  the late-time decays of the heavy
moduli~\cite{heavy-moduli,Moroi-Randall} and the evaporation of cosmological
defects~\cite{Non-thermal-WIMP}.
Here, we propose another way to generate WDM via the
late-time decays of Q-balls. 

To obtain the sufficient free streaming length needed to suppress the
small-scale perturbations, the current velocity of the LSP should
satisfy~\cite{Non-thermal-WIMP}:
\begin{equation}
v_{0}\simeq 10^{-(7-8)}\;.
\label{wdm-1}
\end{equation}
If neutralinos are produced via late-time decays of Q-balls, the current
velocity of the neutralinos is given by
\begin{equation}
v_{0}=\frac{T_{0}}{T_{d}}\frac{m_{\phi}}{m_{\chi}}\simeq
2.4 \times10^{-8}\left(\frac{1\MEV}{T_{d}}\right)
\left(\frac{m_{\phi}}{10\TEV}\right)\left(\frac{100\GEV}{m_{\chi}}\right)\;,
\label{wdm-2}
\end{equation}
where $T_{0}=2.7\;{\rm K}$ is the current temperature of the cosmic microwave
background and $T_{d}$ is the decay temperature of Q-balls;
$m_{\phi}$ is the mass of the relevant flat direction field $\phi$.
Here, we have assumed that there is no substantial energy
reduction by scattering with thermal backgrounds, which is justified
if the condition in Eq.~(\ref{wdm-0}) is satisfied.

Therefore the resultant neutralinos
from late-time decays of Q-balls may serve as a WDM,
for example,
in ``focus point supersymmetry''  with $m_{0}={\cal O}(10\TEV)$ or in
``effective SUSY models'' with very large soft masses for the squarks
in the first and second families $m_{\widetilde{Q}_{1,2}}^2={\cal
O}(10\TEV)^2$~\cite{effective-SUSY}.

\section{Conclusions and discussions}
\label{conclusions}
Both of the production mechanisms of the observed baryon asymmetry
and natures of dark matter are among the most fundamental
problems in particle physics as well as in cosmology.
In the SUSY framework, there exist rather stringent upper bounds on
reheating temperatures of inflation because of the ``cosmological
gravitino problem'', which strongly constrains various
baryo/leptogenesis scenarios requiring relatively high reheating
temperatures.
Affleck--Dine baryogenesis is one of the most
promising candidate to generate the observed baryon asymmetry
in low reheating temperatures, which is free from the
gravitino problem.

In recent developments, it has become clear that the formation of
Q-balls and their late-time decays are almost inevitable consequences
of AD baryogenesis.\footnote{This is not the case for the AD
  leptogenesis via the $LH_u$ flat direction. See Sec. II.} This
requires a significantly large annihilation cross section of the LSP
for the resultant LSPs not to overclose the universe.

In this paper, we investigated in detail the predictions of
the late-time decays of Q-balls.
First, we reviewed the AD
baryogenesis, the Q-ball formation and decays, with thermal
effects taken into account.
Second, we discussed the non-thermal
production of dark matter via the late-time decays of Q-balls with and
without entropy production.
Finally, we specified the allowed
parameter space where the neutralinos produced via the decays of
Q-balls result in a desired mass density of the dark matter in several
SUSY breaking models, and then discussed the prospects of direct and
indirect detection of these non-thermal dark matter candidates.
We also discussed the possibility of WDM generation
via Q-ball decays.

As an interesting new twist, we pointed out that the
AD baryogenesis without superpotential is now one of the
most beautiful scenarios.  Quite a reasonable set of
parameters can explain the observed baryon asymmetry
without any assumptions on additional late-time
entropy productions, such as the decay of heavy moduli
or a thermal inflation.
The resultant baryon asymmetry and dark matter density
are determined solely by the potential for the AD field
and the annihilation cross section of the LSP.
They are completely independent
of the reheating temperature of the inflation and any
other details in the history of the universe.
\\

The parameter space and detection possibility for each
SUSY breaking model can be summarized as follows:
\\

$\bullet$ In minimal supergravity scenarios, allowed parameter space
appears in the ``focus-point'' region where the dominant component of
the LSP is provided by $\widetilde{H}$. A large content of $\widetilde{H}$ in the
LSP strongly enhances the direct detection rate through the Higgs exchange
diagrams. As a result, the direct detection rate is within the reach
of various next-generation detectors in most of the parameter space.
The detection rate of the monoenergetic $\gamma$-rays via 
neutralino annihilation is also enhanced with respect to  bino-like dark
matter, especially in pure $\widetilde{H}$ LSP regions.  Somewhat cuspy
density profiles of the halo $j\gsim 10^{3.5}$ allow us to observe
the monoenergetic $\gamma$-ray lines in the case of pure $\widetilde{H}$ dark
matter.  \\

$\bullet$ In anomaly-mediated SUSY breaking models, the LSP is
$\widetilde{W}$-like in most of the parameter space, which is now a promising
candidate of the dark matter because of the late-time decays of
Q-balls.  The direct detection rate highly depends on ${\rm tan}\beta$
and becomes very large in the case of large ${\rm tan}\beta$. This is
because  the coupling of the heavy Higgs boson with the nucleon is
proportional to ${\rm tan}\beta$, and also the mass of the heavy Higgs
boson is strongly reduced by the large down-type Yukawa couplings.
The indirect detection rate using monoenergetic $\gamma$-rays is
surprisingly large, as reported in Ref.~\cite{AMSB-photon}, and is almost
independent of ${\rm tan}\beta$. Even if we do not assume a highly
cuspy density profile, the detection rate is within the reach of the next
generation of air Cherenkov telescopes~\cite{VERITAS,HESS,MAGIC}.  \\

$\bullet$ In no-scale models with non-universal gaugino masses, the
LSP is $\widetilde{W}$-like in a wide parameter space. (The region
with the $\widetilde{H}$ LSP is excluded by the Higgs mass bound and the
constraints on the $b\rightarrow s\gamma$ branching ratio.)  The
direct and indirect detection rates are comparable with those in
anomaly-mediation models, and within the reach of various
next-generation detectors.
\\

Although we have concentrated on the monoenergetic $\gamma$-rays for
the indirect detection in this paper, the possibility of detecting
the neutralino in other indirect dark matter searches is also
significantly enhanced in the case of $\widetilde{H}$ and $\widetilde{W}$ dark matter.
The prospects of other indirect detections in mSUGRA scenarios
can be obtained in Refs.~\cite{prospects-indirect}.  As for the
anomaly-mediation models, see Refs.~\cite{AMSB-photon,Moroi-Randall}.
Last year, the recent HEAT experiment~\cite{HEAT-new} has confirmed an excess
of a high energy positron flux in cosmic rays~\cite{HEAT-old}.
If the LSP is $\widetilde{H}$- or $\widetilde{W}$-like, the annihilation
into a pair of W-bosons always dominates as long as $m_{\chi}>m_{W}$.
The subsequent decay of a W-boson into a positron, $W^{+}\rightarrow
e^{+}+\nu$, will produce a large excess of positron flux
at and below an energy of about half the W-boson mass.
In fact, it was shown in Ref.~\cite{KANE} that the observed positron
spectrum is naturally explained by the annihilation of
neutralino dark matter, if it is $\widetilde{H}$- or $\widetilde{W}$-like
non-thermal dark matter.  This fact may indicate the existence of
non-thermal sources for these LSPs in the history of the universe.

Finally, we comment on some
advantages of our scenario relative to the
others that can produce the $\widetilde{H}$ or 
$\widetilde{W}$-like dark matter. 
Actually, there exist several other mechanisms that can generate 
the $\widetilde{H}$- or $\widetilde{W}$-like non-thermal dark matter,
which are the decays of heavy moduli fields and/or  heavy gravitinos
with a mass of the order of $100\TEV$~\cite{Moroi-Randall}, and the
evaporations of topological defects~\cite{Non-thermal-WIMP}. 
The former case only appears in the anomaly-mediated SUSY breaking
scenarios. Furthermore, the decays of heavy moduli fields 
are usually accompanied by an extra large entropy production,
which causes a serious problem to explain the observed baryon
asymmetry. If one relies on the AD baryogenesis to produce enough 
baryon asymmetry, then the resultant late-time Q-ball decays 
also play a role as the non-thermal source for the $\widetilde{W}$-like
dark matter. As for the evaporation of topological defects, we 
have no clear reason to expect that they take place at the 
appropriate temperature of $1\MEV\lsim T_{d}\lsim ({\mbox{a few}})\GEV$,
and that they produce the enough initial abundance of the 
$\widetilde{H}$- or the  $\widetilde{W}$-like LSP to be a
dominant component of dark matter.
They may also cause a serious problem by diluting the 
baryon asymmetry.   

In our scenario, the reasonable set of parameters in the scalar
potential of the AD field to explain the 
observed baryon asymmetry naturally leads to the desirable decay
temperature of the Q-balls for the non-thermal production 
of these LSPs in various SUSY breaking scenarios. 
The baryon asymmetry is, of course,  guaranteed by the 
AD mechanism.  
\\

As we have seen in this article, $\widetilde{H}$- and
$\widetilde{W}$-like
LSPs are now promising candidates for the dark matter.
The possibility of direct and indirect detection of these 
dark matter candidates is much larger than that of the standard bino-like LSP.
If $\widetilde{H}$ or $\widetilde{W}$ dark matter is indeed confirmed 
in future experiments, it may shed a bright light on the 
origin of the whole matter in our universe.
\\

Finally, we add a brief comment on the fine tunings on the soft SUSY
breaking parameters to explain the mass density of dark matter for a
given decay temperature of a Q-ball.  As we have discussed in the
text, the dominant annihilation channel of the higgsino- and wino-like
LSP is given by the decay channel into a pair of gauge bosons via the
lightest chargino exchange. This means that the annihilation cross
section of these LSPs are rather stable under the variation of other
soft parameters, such as slepton and squark masses. (Note that the
variation of the squark masses has only a mild effect on the size of
$\mu$-term in the focus point region.)  Consequently, no strong fine
tuning on the soft SUSY breaking parameters is required for a given
decay temperature of a Q-ball. This is a clear contrast to the
standard pure bino dark matter, which needs a strong degeneracy
between the bino and the lightest stau.

\subsection*{Acknowledgements}
M.F. and K.H. thank the Japan Society for the Promotion of Science 
for the financial support.
\normalsize

\clearpage
%%%%%%%%%%%%%%%%%%%%%%%%%%%%%%%%%%%%%%%%%%%%%%%%%%%%%%%%%%%%%%%%%%%
%%%%%%%%%%%%%%%%%%%%%%%%%%%%%%%%%%%%%%%%%%%%%%
%%%%%%%%%%%%%%%%%%%%%%%%%%%%%%%%%%%%%%%%%%%%%%%%%%%%%%%%%%%%%%%%%%%

%%%%%%%%%%%%%%%%%%%%%%%%%%%%%%%%%%%%%%%%%%%%%%%%%%%%%%%%%%%%
%%%%%%%%%%%%%%%%%%%%%%%%%%%%%%%%%%%%%%%%%%%%%%%%
%%%%%%%%%%%%%%%%%%%%%%%%%%%%%%%%%%%%%%%%%%%%%%%%%%%%%%%%%%%%

%%%%%%%%%%%%%%%%%%%%%%%%%%%%%%%%%%%%%%%%%%%%%%%%%%%%%%%%%%%%
%%%%%%%%%%%%%%%%%%%%%%%%%%%%%%%%%%%%%%%%%%%%%%%%
%%%%%%%%%%%%%%%%%%%%%%%%%%%%%%%%%%%%%%%%%%%%%%%%%%%%%%%%%%%%

%%%%%%%%%%%%%%%%%%%%%%%%%%%%%%%%%%%%%%%%%%%%%%%%%%%%%%%%%%%%
%%%%%%%%%%%%%%%%%%%%%%%%%%%%%%%%%%%%%%%%%%%%%%%%
%%%%%%%%%%%%%%%%%%%%%%%%%%%%%%%%%%%%%%%%%%%%%%%%%%%%%%%%%%%%

%%%%%%%%%%%%%%%%%%%%%%%%%%%%%%%%%%%%%%%%%%%%%%%%%%%%%%%%%%%%
%%%%%%%%%%%%%%%%%%%%%%%%%%%%%%%%%%%%%%%%%%%%%%%%
%%%%%%%%%%%%%%%%%%%%%%%%%%%%%%%%%%%%%%%%%%%%%%%%%%%%%%%%%%%%

%%%%%%%%%%%%%%%%%%%%%%%%%%%%%%%%%%%%%%%%%%%%%%%%%%%%%%%%%%%%
%%%%%%%%%%%%%%%%%%%%%%%%%%%%%%%%%%%%%%%%%%%%%%%%
%%%%%%%%%%%%%%%%%%%%%%%%%%%%%%%%%%%%%%%%%%%%%%%%%%%%%%%%%%%%
\clearpage

%%%%%%%%%%%%%%%%%%%%%%%%%%%%%%%%%%%%%%%%%%%%%%%%%%%%%%%%%%%%
%%%%%%%%%%%%%%%%%%%%%%%%%%%%%%%%%%%%%%%%%%%%%%%%
%%%%%%%%%%%%%%%%%%%%%%%%%%%%%%%%%%%%%%%%%%%%%%%%%%%%%%%%%%%%

%%%%%%%%%%%%%%%%%%%%%%%%%%%%%%%%%%%%%%%%%%%%%%%%%%%%%%%%%%%%
%%%%%%%%%%%%%%%%%%%%%%%%%%%%%%%%%%%%%%%%%%%%%%%%
%%%%%%%%%%%%%%%%%%%%%%%%%%%%%%%%%%%%%%%%%%%%%%%%%%%%%%%%%%%%

%%%%%%%%%%%%%%%%%%%%%%%%%%%%%%%%%%%%%%%%%%%%%%%%%%%%%%%%%%%%
%%%%%%%%%%%%%%%%%%%%%%%%%%%%%%%%%%%%%%%%%%%%%%%%
%%%%%%%%%%%%%%%%%%%%%%%%%%%%%%%%%%%%%%%%%%%%%%%%%%%%%%%%%%%%

%%%%%%%%%%%%%%%%%%%%%%%%%%%%%%%%%%%%%%%%%%%%%%%%%%%%%%%%%%%%
%%%%%%%%%%%%%%%%%%%%%%%%%%%%%%%%%%%%%%%%%%%%%%%%
%%%%%%%%%%%%%%%%%%%%%%%%%%%%%%%%%%%%%%%%%%%%%%%%%%%%%%%%%%%%

%%%%%%%%%%%%%%%%%%%%%%%%%%%%%%%%%%%%%%%%%%%%%%%%%%%%%%%%%%%%
%%%%%%%%%%%%%%%%%%%%%%%%%%%%%%%%%%%%%%%%%%%%%%%%
%%%%%%%%%%%%%%%%%%%%%%%%%%%%%%%%%%%%%%%%%%%%%%%%%%%%%%%%%%%%

%%%%%%%%%%%%%%%%%%%%%%%%%%%%%%%%%%%%%%%%%%%%%%%%%%%%%%%%%%%%
%%%%%%%%%%%%%%%%%%%%%%%%%%%%%%%%%%%%%%%%%%%%%%%%
%%%%%%%%%%%%%%%%%%%%%%%%%%%%%%%%%%%%%%%%%%%%%%%%%%%%%%%%%%%%
\vspace{-0.5cm}
%%%%%%%%%%%%%%%%%%%%%%%%%%%%%%%%%%%%%%%%%%%%%%%%%%%%%%%%%%%%
%%%%%%%%%%%%%%%%%%%%%%%%%%%%%%%%%%%%%%%%%%%%%%%%
%%%%%%%%%%%%%%%%%%%%%%%%%%%%%%%%%%%%%%%%%%%%%%%%%%%%%%%%%%%%

%%%%%%%%%%%%%%%%%%%%%%%%%%%%%%%%%%%%%%%%%%%%%%%%%%%%%%%%%%%%
%%%%%%%%%%%%%%%%%%%%%%%%%%%%%%%%%%%%%%%%%%%%%%%%
%%%%%%%%%%%%%%%%%%%%%%%%%%%%%%%%%%%%%%%%%%%%%%%%%%%%%%%%%%%%
\vspace{-0.5cm}
%%%%%%%%%%%%%%%%%%%%%%%%%%%%%%%%%%%%%%%%%%%%%%%%%%%%%%%%%%%%
%%%%%%%%%%%%%%%%%%%%%%%%%%%%%%%%%%%%%%%%%%%%%%%%
%%%%%%%%%%%%%%%%%%%%%%%%%%%%%%%%%%%%%%%%%%%%%%%%%%%%%%%%%%%%

%%%%%%%%%%%%%%%%%%%%%%%%%%%%%%%%%%%%%%%%%%%%%%%%%%%%%%%%%%%%
%%%%%%%%%%%%%%%%%%%%%%%%%%%%%%%%%%%%%%%%%%%%%%%%
%%%%%%%%%%%%%%%%%%%%%%%%%%%%%%%%%%%%%%%%%%%%%%%%%%%%%%%%%%%%
\vspace{-0.5cm}
%%%%%%%%%%%%%%%%%%%%%%%%%%%%%%%%%%%%%%%%%%%%%%%%%%%%%%%%%%%%
%%%%%%%%%%%%%%%%%%%%%%%%%%%%%%%%%%%%%%%%%%%%%%%%
%%%%%%%%%%%%%%%%%%%%%%%%%%%%%%%%%%%%%%%%%%%%%%%%%%%%%%%%%%%%

%%%%%%%%%%%%%%%%%%%%%%%%%%%%%%%%%%%%%%%%%%%%%%%%%%%%%%%%%%%%
%%%%%%%%%%%%%%%%%%%%%%%%%%%%%%%%%%%%%%%%%%%%%%%%
%%%%%%%%%%%%%%%%%%%%%%%%%%%%%%%%%%%%%%%%%%%%%%%%%%%%%%%%%%%%
\vspace{-0.5cm}
%%%%%%%%%%%%%%%%%%%%%%%%%%%%%%%%%%%%%%%%%%%%%%%%%%%%%%%%%%%%
%%%%%%%%%%%%%%%%%%%%%%%%%%%%%%%%%%%%%%%%%%%%%%%%
%%%%%%%%%%%%%%%%%%%%%%%%%%%%%%%%%%%%%%%%%%%%%%%%%%%%%%%%%%%%

\end{document}